\documentclass[sigconf,screen]{acmart}

\usepackage{color}
\usepackage{bbm}
\usepackage{multirow}
\usepackage[inline]{enumitem}
\usepackage{graphicx}
\usepackage{subcaption}
\usepackage{sistyle}
\usepackage[ruled]{algorithm2e}
\usepackage{booktabs}
\usepackage{caption}
\SIthousandsep{,}
\usepackage{makecell}
\usepackage{amsmath}
\usepackage[english]{babel}
\usepackage{hyperref}
\usepackage{array} 
\usepackage{graphicx}
\usepackage{algorithmic}
\usepackage{tikz}
\usepackage{wrapfig}
\usepackage{acronym}
\usepackage[utf8]{inputenc}
\usepackage{CJKutf8}
\usepackage{xcolor}
\newcommand{\heading}[1]{\vspace*{.5mm}\noindent\textbf{#1.}}

\AtBeginDocument{%
  \providecommand\BibTeX{{%
    \normalfont B\kern-0.5em{\scshape i\kern-0.25em b}\kern-0.8em\TeX}}}

\definecolor{lightred}{rgb}{1, 0.8, 0.8}

\makeatletter
\g@addto@macro\normalsize{%
  \abovedisplayskip 3pt plus1pt 
  \belowdisplayskip 3pt plus1pt
  \abovedisplayshortskip  0pt plus1pt%
  \belowdisplayshortskip  0pt plus1pt
}
\makeatother

\acrodef{CV}{computer vision}
\acrodef{IR}{information retrieval}
\acrodef{LLM}{large language model}
\acrodef{MDP}{Markov decision process}
\acrodef{NLP}{natural language processing}
\acrodef{NRM}{neural ranking model}
\acrodef{RL}{reinforcement learning}

\author{Hongru Song}
\author{Yu-An Liu}
\affiliation{
 \institution{State Key Laboratory of AI Safety, Institute of Computing Technology, Chinese Academy of Sciences}
 \institution{University of Chinese Academy of Sciences}
 \city{Beijing}
 \country{China}
}
\email{{songhongru24s, liuyuan21b}@ict.ac.cn}

\author{Ruqing Zhang}
\author{Jiafeng Guo}
\affiliation{
 \institution{State Key Laboratory of AI Safety, Institute of Computing Technology, Chinese Academy of Sciences}
 \institution{University of Chinese Academy of Sciences}
 \city{Beijing}
 \country{China}
}
\email{{zhangruqing,guojiafeng}@ict.ac.cn}

\author{Maarten de Rijke}
\affiliation{
 \institution{University of Amsterdam}
 \city{Amsterdam}
 \country{The Netherlands}
}
\email{m.derijke@uva.nl}

\author{Sen Li}
\author{Wenjun Peng}

\affiliation{
 \institution{Researcher}
 \city{Hangzhou}
 \country{China}
}
\email{lisen.lisen@alibaba-inc.com}
\email{pengwj@mail.ustc.edu.cn}

\author{Fuyu Lv}
\affiliation{
 \institution{Researcher}
 \city{Hangzhou}
 \country{China}
}
\email{fuyu.lfy@alibaba-inc.com}

\author{Xueqi Cheng}
\affiliation{
 \institution{State Key Laboratory of AI Safety, Institute of Computing Technology, Chinese Academy of Sciences}
 \institution{University of Chinese Academy of Sciences}
 \city{Beijing}
 \country{China}
}
\email{cxq@ict.ac.cn}
\renewcommand{\shortauthors}{Song et al.}

\setcopyright{acmlicensed}
\copyrightyear{2018}
\acmYear{2018}
\acmDOI{XXXXXXX.XXXXXXX}
\acmConference[Conference acronym 'XX]{Make sure to enter the correct
  conference title from your rights confirmation email}{June 03--05,
  2018}{Woodstock, NY}
\acmISBN{978-1-4503-XXXX-X/2018/06}

\makeatother


\vspace{-2mm}

\begin{document}

\title[LLMs as Sparse Retrievers: A Framework for First-Stage Product Search]{LLMs as Sparse Retrievers:\\ A Framework for First-Stage Product Search}

\begin{abstract}
Product search is a crucial component of modern e-commerce platforms, with billions of user queries every day. In product search systems, first-stage retrieval should achieve high recall while ensuring efficient online deployment. 
Sparse retrieval is particularly attractive in this context due to its interpretability and storage efficiency. 
However, sparse retrieval methods suffer from severe vocabulary mismatch issues, leading to suboptimal performance in product search scenarios.

With their potential for semantic analysis, large language models (LLMs) offer a promising avenue for mitigating vocabulary mismatch issues and thereby improving retrieval quality. Directly applying LLMs to sparse retrieval in product search exposes two key challenges:  
\begin{enumerate*}[label=(\roman*)] 
\item Queries and product titles are typically short and highly susceptible to LLM-induced hallucinations, such as generating irrelevant expansion terms or underweighting critical literal terms like brand names and model numbers;
\item The large vocabulary space of LLMs leads to difficulty in initializing training effectively, making it challenging to learn meaningful sparse representations in such ultra-high-dimensional spaces. 
\end{enumerate*} 
To address these challenges, we propose PROSPER, a framework for PROduct search leveraging LLMs as SParsE Retrievers. PROSPER incorporates: 
\begin{enumerate*}[label=(\roman*)] 
\item A literal residual network that alleviates hallucination in lexical expansion by reinforcing underweighted literal terms through a residual compensation mechanism; and
\item A lexical focusing window that facilitates effective training initialization via a coarse-to-fine sparsification strategy. 
\end{enumerate*}
Extensive offline and online experiments show that PROSPER significantly outperforms sparse baselines and achieves recall performance comparable to advanced dense retrievers, while also achieving revenue increments online.
\end{abstract}
\maketitle

\vspace{-2mm}
\section{Introduction}
E-commerce platforms have become an integral part of daily life. For many online consumers, product search engines serve as both the entry point and the central hub that connects them to a vast array of products. 
The primary challenge lies in efficiently handling search queries from hundreds of millions of users over a billion-scale product catalog, all under strict latency constraints.  
To meet these demands, industrial-scale search engines generally follow the paradigm of ``index-retrieve-then-rank'' \cite{10.1145/3637528.3671654}. Here, first-stage retrieval is very critical, as it determines the recall ceiling, i.e., any relevant products missed here cannot be recovered in later stages.

\heading{Dense retrieval: The prevailing paradigm and its challenges}
Dense retrieval uses neural networks to encode queries and documents into low-dimensional dense vectors, enabling semantic matching at scale \cite{guoSemanticModelsFirststage2022,karpukhin-etal-2020-dense, 10.1145/3397271.3401075}. 
However, it poses significant challenges for industrial-scale e-commerce applications:
\begin{enumerate*}[label=(\roman*)]
\item Dense vectors have a ``black-box'' nature, lacking interpretability and making it difficult to understand the model's decisions;
\item They introduce substantial indexing and storage overheads. For example, on the MS MARCO passage ranking dataset \cite{bajaj2018msmarcohumangenerated}, dense retrieval methods require index sizes that are several to dozens of times larger than the original corpus \cite{10.1145/3626772.3657951,xuCSPLADELearnedSparse2025}. 
\end{enumerate*}

\heading{Sparse retrieval: A return to interpretability and storage efficiency}
In contrast to dense retrieval, sparse retrieval operates on high-dimensional but sparse vectors, where each dimension corresponds to a specific term in the vocabulary, and the value indicates the importance of that term \cite{9357643,10.1561/1500000019,10.1145/3404835.3463098}. 
Owing to its inherent advantages in interpretability and storage efficiency, sparse retrieval has remained a key component in industrial-scale search systems.  
The development of sparse retrieval has evolved from classical statistical models \cite{9357643,10.1561/1500000019} to learned sparse retrieval \cite{10.1145/3269206.3271800,10.1145/3397271.3401204,10.1145/3404835.3463098}. 

A major milestone is SPLADE \cite{10.1145/3404835.3463098}, which uses BERT to jointly learn term expansion and weighting, significantly improving retrieval effectiveness.  
However, SPLADE is inherently limited by the pre-trained knowledge and semantic capacity of its BERT backbone, which constrains its ability to address the fundamental problem of vocabulary mismatch. Moreover, as shown in our preliminary analysis (see Section~\ref{subsec:Application and Challenges}), even strong models like SPLADE exhibit a noticeable performance gap compared to classic dense retrievers when directly applied to product search scenarios.

\heading{LLMs as sparse retrievers: Challenges}
Large language models (LLMs), with their pre-trained knowledge and potential for semantic analysis \cite{qwen2025qwen25technicalreport,taoLLMsAreAlso2024,minaee2025largelanguagemodelssurvey}, present a promising opportunity to advance product search by balancing semantic understanding with interpretability and storage efficiency. 
We explore using LLMs as a backbone for sparse retrieval, aiming to leverage their semantic strength while retaining the advantages of sparse methods. 
However, preliminary results (Section~\ref{subsec:Application and Challenges}) show that simply replacing BERT with an LLM like Qwen2.5-3B \cite{qwen2025qwen25technicalreport} in SPLADE introduces two key issues: 
\begin{enumerate*}[label=(\roman*)]
\item \emph{Lexical expansion hallucination}. The model tends to overemphasize expanded or even irrelevant terms, while underweighting literal terms that are essential for capturing user intent, such as brand and model names. 
\item \emph{Unstable training initialization}. Due to the large vocabulary space of LLMs, training becomes significantly more difficult without proper guidance. The model must learn to expand short user queries and product titles within an ultra-high-dimensional space (e.g., exceeding 15,000 dimensions in Qwen2.5), which hampers learning stability and efficiency.
\end{enumerate*}

\heading{LLMs as sparse retrievers: Solutions} 
To address the issues outlined above, we propose PROSPER, a framework for PROduct search using LLMs as SParsE Retrievers. PROSPER introduces contributions at both the architectural and training levels: 
\begin{enumerate*}[label=(\roman*)]
\item We design a \emph{literal residual network} (LRN) that employs a compensatory weighting mechanism to amplify the importance of literal terms in user queries and items. This effectively mitigates hallucination by anchoring the model's attention to user-critical tokens, such as brand and model names. 

\item We introduce a \emph{lexical focusing window} (LFW), which works in concert with FLOPS regularization \cite{paria2020minimizingflopslearnefficient} to guide the model through a coarse-to-fine sparsification process. LFW acts as a hard constraint during early training stages to force the model to rapidly achieve sparsification and escape from ultra-high-dimensional representation learning as quickly as possible, while FLOPS regularization provides fine-grained control as training progresses, ensuring efficient and targeted learning from the outset.
\end{enumerate*}

To the best of our knowledge, this is the first study to explore the use of LLMs for sparse retrieval in product search. 

\heading{Experimental findings}
We evaluate PROSPER through offline experiments on both the public Multi-CPR E-commerce dataset and a real-world dataset sampled from Taobao search logs.
PROSPER achieves a substantial 10.2\% improvement in the key target product recall metric over the BM25 baseline \cite{10.1561/1500000019}, and a 4.3\% gain over the SPLADE baseline \cite{formalSPLADEV2Sparse2021}, while delivering performance on par with advanced dense retrieval models. Furthermore, we conduct online experiments in the Taobao search\footnote{Taobao (\url{https://www.taobao.com/}) is one of China's largest e-commerce platforms.} and achieve a 0.64\% improvement in the key GMV metric. 

\vspace*{-2mm}
\section{Preliminaries}
\label{sec:preliminary}
A formal definition of first-stage product search is provided in Appendix~\ref{app:PB}.
This section introduces the SPLADE framework, which serves as the foundation of our method, and outlines the challenges we encountered when implementing SPLADE in product search. 

\vspace*{-2mm}
\subsection{SPLADE}
\heading{Model architecture} Given an input query or document sequence (after WordPiece tokenization) $S = (t_1, t_2, \ldots, t_N)$, and its corresponding BERT embeddings $\mathbf{h} = (h_1, h_2, \ldots, h_N)$, SPLADE projects each hidden representation to a vocabulary-sized vector $\mathbf{H}_i \in \mathbb{R}^{|V|}$ with the masked language modeling head. The $j$-th dimension of $\mathbf{H}_i$ represents the importance of vocabulary token $j$ to input token $i$, which in practice is the logit from the language modeling head output. The final representation is obtained by applying ReLU activation and log-saturation effect to each token's logits, followed by max-pooling across token positions:
\begin{equation}
    w_j = \max_{i \in t} \log(1 + \text{ReLU}(w_{ij})),
\end{equation}
where $w_{ij}$ represents the importance of vocabulary token $j$ for input token $i$, and $w_j$ is the final weight for term $j$ in the representation.

\heading{Model training} SPLADE is trained using a combination of ranking loss and regularization terms. The ranking loss employs two negative sampling strategies: in-batch negatives~\cite{lin-etal-2021-batch,qu-etal-2021-rocketqa} and hard negative sampling \cite{xiong2020approximatenearestneighbornegative}. Given a query $q_i$ in a batch, a positive document $d_i^+$, a hard negative document $d_i^-$ (e.g., from BM25 sampling), and a set of negative documents in the batch (positive documents from other queries) $\{d_{i,j}^-\}_j$, the model is trained with InfoNCE loss \cite{oord2019representationlearningcontrastivepredictive} for contrastive training:
\begin{equation}
    \mathcal{L}_\mathit{rank\text{-}IBN} = -\log \frac{e^{s(q_i, d_i^+)}}{e^{s(q_i, d_i^+)} + e^{s(q_i, d_i^-)} + \sum_j e^{s(q_i, d_{i,j}^-)}},
\end{equation}
where $s(q, d)$ denotes the ranking score obtained via dot product between query and document representations.

To encourage sparsity in learned representations, SPLADE adopts FLOPS regularization \cite{paria2020minimizingflopslearnefficient}, a smooth approximation of the average floating-point operations required for scoring, thus directly tied to retrieval latency.  
It is defined using $a_j$ as a continuous relaxation of the activation (i.e. the term has a non zero weight) probability $p_j$ for token $j$, and estimated for documents $d$ in a batch of size $\mathcal{N}$ by $\bar{a}_j = \frac{1}{\mathcal{N}} \sum_{i=1}^\mathcal{N} w_j^{(d_i)}$. The FLOPS loss is defined as:
\begin{equation}
    \mathcal{L}_\mathit{FLOPS} = \sum_{j \in V} \bar{a}_j^2 = \sum_{j \in V} \left(\frac{1}{\mathcal{N}} \sum_{i=1}^\mathcal{N} w_j^{(d_i)}\right)^2. 
\end{equation}
This regularization differs from the $\ell_1$ regularization used in SNRM \cite{10.1145/3269206.3271800} as it penalizes high average term weights, promoting a more balanced index distribution. The overall training objective combines ranking loss with regularization terms:
\begin{equation}
\label{eq:splade_loss}
    \mathcal{L} = \mathcal{L}_\mathit{rank\text{-}IBN} + \lambda_q \mathcal{L}_\mathit{FLOPS}^q + \lambda_d \mathcal{L}_\mathit{FLOPS}^d,
\end{equation}
where $\lambda_q$ and $\lambda_d$ are hyperparameters to tune the regularization strength for queries and documents, respectively, allowing different sparsity control for each. 
While later SPLADE variants \cite{10.1145/3477495.3531857,formalSPLADEV2Sparse2021} have introduced additional training techniques such as distillation, our work focuses on contrastive learning, as those more complex methods are orthogonal to the approach. 

\subsection{Exploration and challenges} 
\label{subsec:Application and Challenges}
While SPLADE \cite{10.1145/3404835.3463098,formalSPLADEV2Sparse2021} has demonstrated a performance comparable to dense retrieval on several benchmarks, especially in passage retrieval, its practical effectiveness in product search scenarios remains underexplored. 

\heading{Difference between passage retrieval and product retrieval}
For traditional passage retrieval, passages are \emph{content-rich} with abundant \emph{noisy terms}, requiring the model to identify \emph{important terms} for expansion while discarding \emph{noisy terms}. However, in product search, both user queries and product titles are essentially \emph{keyword aggregations} with high information density, which presents distinct challenges :
\begin{enumerate*}[label=(\roman*)]
\item for a user query, the model must accurately understand the \emph{short and specific} user needs, where all literal query terms are typically important;
\item for a product title, since product titles are \emph{brief}, the model needs to expand representations richly enough to \emph{comprehensively cover potential query terms}. 
\end{enumerate*}

\heading{Implementing SPLADE in product search} In our initial experiments with a BERT-based SPLADE model on the Multi-CPR E-commerce test dataset \cite{10.1145/3477495.3531736}, we achieved a Hit@1000 of 89.6\%. While decent, this lags behind the 92.1\% achieved by DPR and the 96\%+ reported on MS MARCO passage ranking dataset \cite{bajaj2018msmarcohumangenerated}. 

\heading{Enhancing SPLADE with LLMs for product search} 
Recently, researchers have begun adapting LLMs for sparse retrieval (see Appendix~\ref{sec:related_work} for detailed related work). 
\citet{zengScalingSparseDense2025} studied the scaling laws of sparse retrievers based on LLMs, while CSPLADE, following Nv-Embed \cite{leeNVEmbedImprovedTechniques2025} and echo embedding \cite{springerRepetitionImprovesLanguage2024}, solved the information flow issues caused by unidirectional attention when applying LLMs to enhance SPLADE.

Inspired by these works, we first replaced BERT with Qwen-2.5-3B \cite{qwen2025qwen25technicalreport} and this raised the Hit@1000 score to 91.3\% in Multi-CPR E-commerce test dataset \cite{10.1145/3477495.3531736}. 
We also referenced previous work on optimizing attention mechanisms in LLMs to enhance representation quality \cite{behnamghaderLLM2VecLargeLanguage2024,springerRepetitionImprovesLanguage2024,xuCSPLADELearnedSparse2025}, achieving performance improvements up to 92.4\%, but the ranking of the target products remained suboptimal. 

\begin{table}[t]
\caption{Examples of lexical expansion hallucination in LLM-based SPLADE. Terms are ranked by weight in descending order. Literal terms are shown in black, useful expansions in green, and noisy expansions in red.}
\label{tab:hallucination}
\centering
\includegraphics[width=0.9\linewidth]{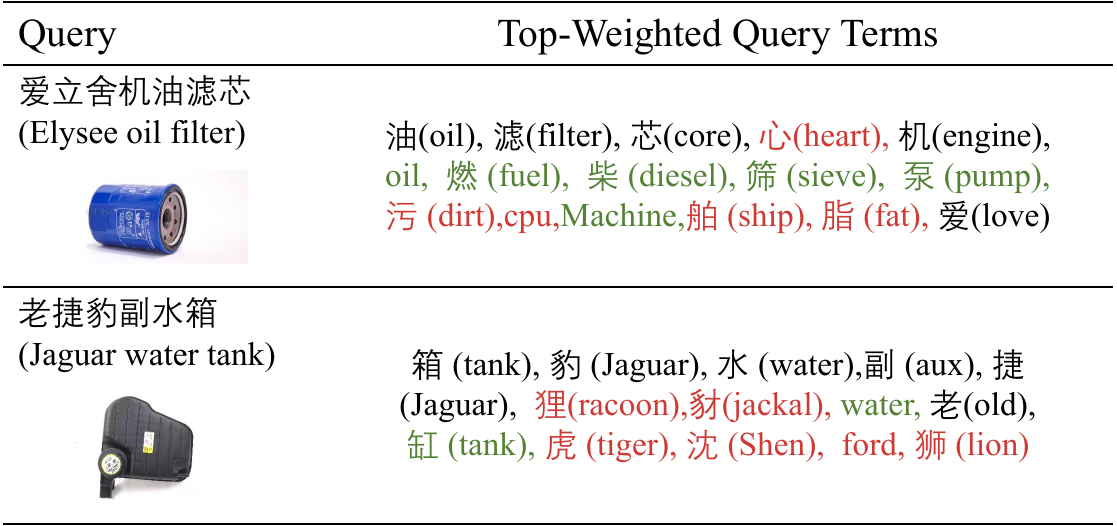}
\end{table}
\heading{Case study and challenges} Case analyses revealed severe \emph{lexical expansion hallucinations} (see Table~\ref{tab:hallucination}):
\begin{enumerate*}[label=(\roman*)]
\item Literal terms like brands remained underweighted (e.g., \begin{CJK}{UTF8}{gbsn}“爱立舍机油滤芯”\end{CJK}, where ``Elysee'' was overlooked);
\item Irrelevant expansions appeared in ambiguous or rare queries (e.g., \begin{CJK}{UTF8}{gbsn}“老捷豹副水箱”\end{CJK}, where the model expanded irrelevant animal terms).
\end{enumerate*}
Moreover, LLMs typically have very large vocabularies. For example, the Qwen-2.5 model \cite{qwen2025qwen25technicalreport} contains over 150,000 dimensions, which is excessive compared to the short length of user queries (averaging 6 terms in the Taobao-Internal dataset) and product titles (averaging 25 terms in the Taobao-Internal dataset). Without proper constraints, sparsifying high-dimensional spaces leads to unstable early training, preventing the model from learning meaningful sparse representations.

\vspace*{-1mm}
\section{Method}
\label{sec:method}
\begin{figure*}[t]
    \centering
    \begin{subfigure}[b]{0.69\textwidth}
        \centering
        \includegraphics[width=\linewidth, trim={0 0 426.49pt 0}, clip]{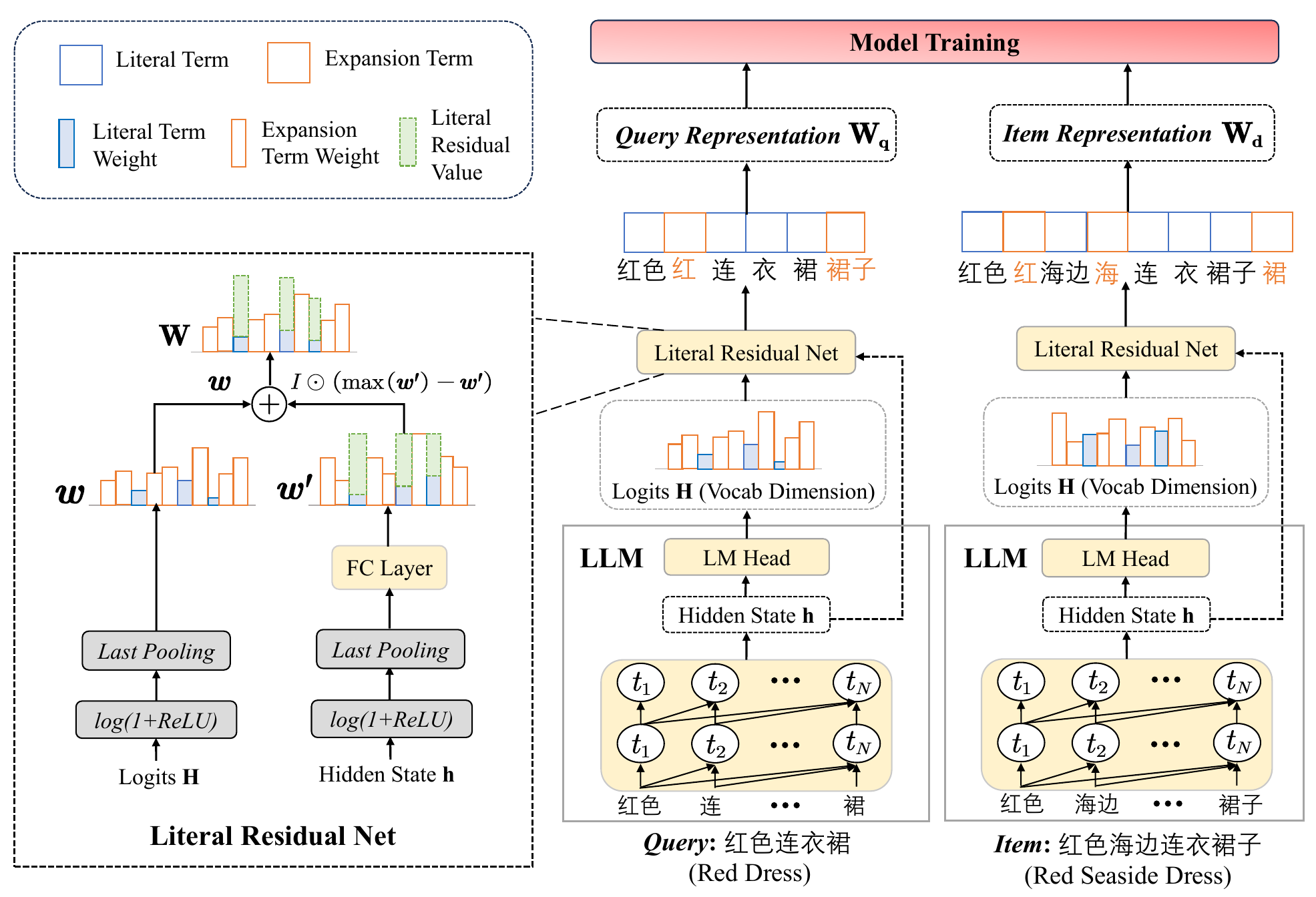}
        \caption{Model architecture}
        \label{fig:model_a}
    \end{subfigure}%
    \hfill
    \begin{subfigure}[b]{0.31\textwidth}
        \centering
        \includegraphics[width=\linewidth, trim={949.29pt 0 0 0}, clip]{figs/Overview.pdf}
        \caption{Model training}
        \label{fig:model_b}
    \end{subfigure}
    \caption{Overview of the proposed PROSPER framework. 
    (a) The model architecture, detailing the query/item encoding process and the core LRN module that enhances literal term weights.
    (b) The model training, illustrating the LFW, the asymmetric similarity computation, and the final loss function combining ranking loss with FLOPS regularization.}
    \label{fig:overall_model}
    \vspace{-4mm}
\end{figure*}

As illustrated in Figure~\ref{fig:overall_model}, the PROSPER framework features innovations in two key areas:
\begin{enumerate*}[label=(\roman*)]
\item model architecture, where we introduce the literal residual network to alleviate lexical expansion hallucination; and
\item model training, where we introduce the lexical focusing window to provide more targeted training initialization. 
\end{enumerate*}
\vspace{-2mm}
\subsection{Model architecture}
\label{sec:model_architecture}
\heading{LLM-based representation} 
Building upon SPLADE \cite{formalSPLADEV2Sparse2021}, our approach uses LLMs as the backbone model to enhance semantic understanding capabilities. For an input query or item sequence (after tokenization) $S = (t_1, t_2, ..., t_N)$, we extract the last-layer hidden states $\mathbf{h} = (h_1, h_2, ..., h_N)$ from the LLM. Each hidden representation is then projected to a vocabulary-sized vector through the language modeling head(LM head), obtaining logits $\mathbf{H}_i \in \mathbb{R}^{|V|}$ for each token position $i$. After extracting the hidden states $\mathbf{h}$ and logits $\mathbf{H}$ of the last layer, we input them into the literal residual network.

\heading{Literal residual network (LRN)}
The core idea of LRN is to guide the LLM to first focus on literal terms before expanding to related terms, assigning sufficient weights to critical terms such as product brands and product models that represent specific user needs.

For the hidden states $\mathbf{h}$ and logits $\mathbf{H}$ extracted from the LLM, we first apply ReLU activation and log-saturation transformation:
\begin{equation}
    \tilde{h}_{i} = \log(1 + \text{ReLU}(h_{i})), ~~~~~~
    \tilde{H}_{ij} = \log(1 + \text{ReLU}(H_{ij})),
\end{equation}
where $h_i$ is the hidden state vector for the $i$-th input token, and $H_{ij}$ represents the importance of the $j$-th vocabulary term for the $i$-th input token. 

Next, we need to pool these representations across the token positions. While SPLADE uses max-pooling \cite{formalSPLADEV2Sparse2021}, this is suboptimal for LLMs. Unlike bidirectional attention in BERT, LLMs employ causal attention where each token can only attend to previous tokens. To mitigate this limitation, we adopt last-pooling \cite{taoLLMsAreAlso2024,behnamghaderLLM2VecLargeLanguage2024}, which takes the representation of the last token position of the input sequence, as it effectively aggregates information from the entire sequence. The basic representation $\mathbf{w}$ is derived by last-pooling the transformed logits $\tilde{H}$:
\begin{equation}
    \mathbf{w}_{j} = \text{Last}(\tilde{H}_{ij}),
\end{equation}
where $\mathbf{w}_{j}$ is the weight for the $j$-th term in the vocabulary.

Meanwhile, we obtain a pooled hidden state representation by last-pooling the transformed hidden states $\tilde{h}$:
\begin{equation}
    \tilde{h}_{\text{last}} = \text{Last}(\tilde{h}).
\end{equation}
Then, the LRN feeds this vector $\tilde{h}_{\text{last}}$ through a fully connected layer $\text{FC}_{\text{layer}}$ that projects from hidden dimensions to vocabulary dimensions, producing the enhancement vector $\mathbf{w}' \in \mathbb{R}^{|V|}$:
\begin{equation}
    \mathbf{w}' = \text{FC}_{\text{layer}}(\tilde{h}_{\text{last}}).
\end{equation}

After obtaining both the basic representation $\mathbf{w}$ and the enhancement vector $\mathbf{w}'$, we aggregate them to derive the final representation $\mathbf{W}$:
\begin{equation}
    \mathbf{W} = \mathbf{w} + \mathbf{I} \odot (\max(\mathbf{w}') - \mathbf{w}'),
\end{equation}
where $\mathbf{I}$ is an indicator vector that takes value 1 at positions corresponding to literal terms in the sequence $S$ and 0 elsewhere, $\odot$ denotes element-wise multiplication, and $\max(\mathbf{w}')$ represents the maximum weight across all dimensions in the $\mathbf{w}'$. The weights $\max(\mathbf{w}') - \mathbf{w}'$ serve as the literal residual value, which is used to enhance the literal term weights. Then we utilize the final representation $\mathbf{W}$ and the basic representation $\mathbf{w}$ to train the model.

\heading{Discussion} The way LRN works is like a compensatory weighting mechanism. For an literal term $t_i$ with basic weight $w_{t_i}$, when $w_{t_i}$ is low, indicating that the model has not sufficiently attended to the literal term $t_i$, we compensate with more weight to make the model focus more on this term. Conversely, if $w_{t_i}$ is already high, indicating that the model has adequately attended to the term, less compensation is provided. Through this flexible literal term weight compensation mechanism, we can guide the model to gradually attend to product brands, product models, and other terms that represent specific user needs.

\subsection{Model training}
\label{sec:model_training}

\heading{Coarse-to-fine sparsification}
Although FLOPS regularization was introduced to control sparsity, it operates as a soft mechanism and is insufficient to effectively guide the model during the critical early stages of training.
We argue that sparsification should follow a coarse-to-fine strategy: early training should enforce rapid sparsification to escape the high-dimensional space efficiently and establish a strong foundation, while later stages should adopt gradual refinement to balance retrieval quality and sparsity.

\heading{Lexical focusing window}
Building on the coarse-to-fine idea, we propose a Lexical Focusing Window (LFW) to guide early training. 
The LFW operates by applying a conditional top-k pooling operator, $\text{TopK}_k$, which applies pooling only when the number of non-zero dimensions in a representation $\mathbf{W}$ exceeds the window size $k$:
\begin{equation}
\label{eq:conditional_topk}
\text{TopK}_k(\mathbf{W}) = 
\begin{cases} 
    \text{TopK}(\mathbf{W}, k) & \text{if } ||\mathbf{W}||_0 > k, \\
    \mathbf{W} & \text{if } ||\mathbf{W}||_0 \le k, 
\end{cases}
\end{equation}
where $||\cdot||_0$ is the $\ell_0$ norm, counting the non-zero elements in the vector, and the standard $\text{TopK}$ function retains the $k$ largest weights while setting others to zero. This conditional application ensures that the LFW acts primarily during the early, dense stages of training and gracefully phases out as the representations become naturally sparse.

The LFW and FLOPS regularization work in synergy:
\begin{enumerate*}[label=(\roman*)]
\item in the early training stages, the LFW acts as a hard constraint, forcing the model to focus on a limited set of the most important terms, enabling rapid and targeted sparsification;
\item once the representation has stabilized, the LFW's role diminishes, and the FLOPS regularizer takes over for fine-grained adjustments.
\end{enumerate*}

\heading{Training objective}
With the LRN architecture and the LFW mechanism defined, we can now formulate the final training objective. Similarly to SPLADE, our training objective is built upon a contrastive ranking loss, which requires a similarity score between query and item representations. While the original SPLADE employs a standard dot product, we introduce $\ell_2$ normalization on the query side, a modification we found in our experiments to improve both sparsification and retrieval performance (see Appendix~\ref{sec:normalization}). 
The similarity score is computed using the LFW operator as follows:
\begin{equation}
\label{eq:tbsplade_similarity}
    s_{LFW}(\mathbf{W}_q, \mathbf{W}_d) = \frac{\text{TopK}_{k_q}(\mathbf{W}_q)}{||\text{TopK}_{k_q}(\mathbf{W}_q)||_2} \cdot \text{TopK}_{k_d}(\mathbf{W}_d),
\end{equation}
where $\mathbf{W}_q$ and $\mathbf{W}_d$ are the query and item representations from the LRN output, $k_q$ and $k_d$ are their respective LFW sizes, and $||\cdot||_2$ is the $\ell_2$ norm. 
Our analysis shows that dynamic shrinking of the window size did not yield significant improvements, confirming that its main benefit is enabling a robust and efficient training start. 

Our training objective uses an InfoNCE loss function \cite{oord2019representationlearningcontrastivepredictive} that incorporates in-batch negatives \cite{lin-etal-2021-batch}. For a given query representation $\mathbf{W}_{q_i}$, its positive product representation $\mathbf{W}_{d_i^+}$, and the set of in-batch negatives $\{\mathbf{W}_{d_{i,j}^-}\}_j$ (positive documents from other queries in the batch), the ranking loss is:
\begin{equation}
\label{eq:infonce_loss}
    \mathcal{L}_\mathit{rank\text{-}LFW} = -\log \frac{e^{s_{LFW}(\mathbf{W}_{q_i}, \mathbf{W}_{d_i^+})}}{e^{s_{LFW}(\mathbf{W}_{q_i}, \mathbf{W}_{d_i^+})} + \sum_j e^{s_{LFW}(\mathbf{W}_{q_i}, \mathbf{W}_{d_{i,j}^-})}}.
\end{equation}
The overall training objective combines this ranking loss with the FLOPS regularization terms \cite{paria2020minimizingflopslearnefficient} for both queries and items:
\begin{equation}
\label{eq:tbsplade_loss}
    \mathcal{L}' = \mathcal{L}_\mathit{rank\text{-}LFW} + \lambda_q \mathcal{L}_{\mathit{FLOPS}}^q + \lambda_d \mathcal{L}_{\mathit{FLOPS}}^d,
\end{equation}
where $\lambda_q$ and $\lambda_d$ are hyperparameters that balance the regularization of FLOPS. Here, FLOPS regularization is applied to the basic representation $\mathbf{w_q}$ and $\mathbf{w_d}$.

\vspace*{-2mm}
\section{Offline Experiments}
\label{sec:experiment}
In this section, we conduct a series of offline experiments in product search scenario to comprehensively evaluate PROSPER.
\vspace*{-2mm}
\subsection{Experimental setup}
\heading{\emph{Datasets}} We conduct experiments on two datasets that represent both public benchmarks and real-world industrial settings:
\begin{enumerate*}[label=(\roman*)]
\item \textbf{Multi-CPR E-commerce} \cite{10.1145/3477495.3531736}.Multi-CPR is a publicly available, multi-domain Chinese passage retrieval dataset and we utilize its E-commerce subset;
\item \textbf{Taobao-Internal}. To further validate our approach in a real-world industrial setting, we construct a new dataset by sampling approximately 1.07 million query-item pairs from the real user click logs of Taobao Search in June 2025.
\end{enumerate*}

Multi-CPR E-commerce dataset is manually annotated, providing a more direct relevance signal, whereas the Taobao-Internal dataset is constructed from user click logs, where the relevance signals between queries and items are more complex and diverse.
More detailed dataset information such as the length and number of queries or items is shown in the appendix ~\ref{sec:appendix_dataset_details}.

\heading{\emph{Baselines}}
For dense baselines, we compare against: 
\begin{enumerate*}[label=(\roman*)]
\item \textbf{DPR} \cite{karpukhin-etal-2020-dense}, classic dense retrieval baseline.
\item \textbf{BGE Series} \cite{bge_embedding}, a series of powerful dense text embedding models,and we use \textbf{bge-large-zh-v1.5} and \textbf{bge-base-zh-v1.5} for comparison.
\end{enumerate*} 
For sparse baselines, we use: 
\begin{enumerate*}[label=(\roman*)]
\item \textbf{BM25} \cite{10.1561/1500000019}, the classic sparse method. We compare with \textbf{BM25}\textsubscript{BERT} and \textbf{BM25}\textsubscript{Qwen}, which use tokenizers of BERT-base-chinese and Qwen2.5-3B, respectively. 
\item \textbf{Doc2Query} \cite{nogueira2019documentexpansionqueryprediction}, which alleviates vocabulary mismatch by using a seq2seq model to generate potential queries for document. 
\item \textbf{DeepCT}
\cite{dai2019contextawaresentencepassagetermimportance,10.1145/3397271.3401204}, which leverages BERT to evaluate literal term importance. 
\item \textbf{SPLADE} \cite{10.1145/3404835.3463098,formalSPLADEV2Sparse2021}, our main baseline. We use both \textbf{SPLADE} \cite{10.1145/3404835.3463098} and \textbf{SPLADE-v2} \cite{formalSPLADEV2Sparse2021} and adjust their training process for a fair comparison.
\item We implement \textbf{SP}\textsubscript{Qwen-backbone} \cite{zengScalingSparseDense2025} by replacing the BERT backbone with Qwen2.5-3B model. In addition, following previous work \cite{xuCSPLADELearnedSparse2025,springerRepetitionImprovesLanguage2024,behnamghaderLLM2VecLargeLanguage2024,leeNVEmbedImprovedTechniques2025}, we explore optimizing the attention mechanism of LLMs to improve the representation quality:
\textbf{SP}\textsubscript{Qwen-echoembedding}  \cite{springerRepetitionImprovesLanguage2024,xuCSPLADELearnedSparse2025} with duplicated input sequences, and \textbf{SP}\textsubscript{Qwen-bidattention} \cite{behnamghaderLLM2VecLargeLanguage2024,xuCSPLADELearnedSparse2025,leeNVEmbedImprovedTechniques2025} with bidirectional attention mechanism.
\end{enumerate*}

\heading{\emph{Model variants}} 
To validate the effectiveness of our proposed components, we create several variants of PROSPER.

To analyze the core model design, we implement the following variants: 
\begin{enumerate*}[label=(\roman*)]
\item \textbf{PROSPER}\textsubscript{BERT}, which replaces the Qwen2.5-3B \cite{qwen2025qwen25technicalreport} backbone with BERT-base-chinese \cite{devlin-etal-2019-bert};
\item \textbf{PROSPER}\textsubscript{max-pooling}, which uses the max-pooling strategy from SPLADE;
\item \textbf{PROSPER}\textsubscript{bid-attention}, which uses bidirectional instead of causal attention \cite{behnamghaderLLM2VecLargeLanguage2024,xuCSPLADELearnedSparse2025};
and \item \textbf{PROSPER}\textsubscript{echo-emb}, which duplicates the input sequence to simulate a bidirectional receptive field \cite{springerRepetitionImprovesLanguage2024,xuCSPLADELearnedSparse2025,leeNVEmbedImprovedTechniques2025}.
\end{enumerate*}

To investigate the role of the LRN, the variants are: 
\begin{enumerate*}[label=(\roman*)]
\item \textbf{PROSPER}\textsubscript{w/o-LRN}, which removes the LRN module entirely;
\item \textbf{PROSPER}\textsubscript{LRN-add}, which replaces the residual connection with a direct addition ($\mathbf{W} = \mathbf{w} + \mathbf{I} \odot \mathbf{w}'$);
and \item variants that apply LRN only to queries (\textbf{PROSPER}\textsubscript{LRN-q}) or items (\textbf{PROSPER}\textsubscript{LRN-d}).
\end{enumerate*}

To evaluate the LFW, we experiment with the following: \begin{enumerate*}[label=(\roman*)]
\item removing it (\textbf{PROSPER}\textsubscript{w/o-LFW});
\item using a dynamic window size (\textbf{PROSPER}\textsubscript{LFW-dynamic});
and \item using various fixed window sizes for query ($k_q$) and item ($k_d$).
\end{enumerate*}

To study the contributions of literal and expansion terms, we test: 
\begin{enumerate*}[label=(\roman*)]
\item \textbf{PROSPER}\textsubscript{literal} and \textbf{PROSPER}\textsubscript{expansion}, which are trained and evaluated using only literal or expansion terms, respectively, to understand their individual contributions;
and \item \textbf{PROSPER}\textsubscript{mask-lit} and \textbf{PROSPER}\textsubscript{mask-expan}, where the model is trained normally, but either literal or expansion terms are masked out during match and evaluation stage to analyze their roles in a well-trained model.
\end{enumerate*}

\begin{table*}[t]
\centering
\caption{Main results(\%) on the Multi-CPR E-commerce and Taobao-Internal datasets. Best results are in bold. “–” denotes metrics not reported in the original Multi-CPR paper \cite{10.1145/3477495.3531736}.}
\label{tab:main_results}
\setlength{\tabcolsep}{4.0pt}
\begin{tabular}{l l ccccc c ccc}
\toprule
& \multirow{2}{*}{\textbf{Method}} & \multicolumn{5}{c}{\textbf{Multi-CPR E-commerce}} & & \multicolumn{3}{c}{\textbf{Taobao Internal}} \\
\cmidrule(lr){3-7} \cmidrule(lr){9-11}
& & Hit@1 & Hit@10 & Hit@100 & Hit@1000 & MRR@10 & & Recall@10 & Recall@100 & Recall@1000 \\
\midrule
\multirow{3}{*}[-1pt]{\textit{Dense Baselines}}
& DPR & - & - & - & 92.1 & 27.04 & & 42.50 & 72.61 & 91.07 \\
& BGE-base-zh-v1.5 & \textbf{26.6} & 54.1 & 80.4 & 93.7 & 34.70 & & 50.83 & 75.87 & 92.93 \\
& BGE-large-zh-v1.5 & 26.1 & \textbf{55.2} & \textbf{81.9} & 93.1 & \textbf{35.10} & & 50.66 & 76.89 & 93.45 \\
\midrule
\multirow{9}{*}[-1pt]{\textit{Sparse Baselines}}
& BM25\textsubscript{Qwen} & 14.3 & 37.1 & 62.6 & 83.7 & 20.60 & & 39.80 & 65.53 & 84.68 \\
& BM25\textsubscript{BERT} & 16.5 & 40.9 & 66.9 & 85.5 & 23.48 & & 41.23 & 67.84 & 86.21 \\
& Doc2Query & - & - & - & 82.6 & 23.85 & & 42.06 & 68.41 & 86.72 \\
& DeepCT & 21.9 & 46.4 & 72.2 & 87.5 & 29.14 & & 46.64 & 73.26 & 86.90 \\
& SPLADE & 17.2 & 42.7 & 70.6 & 89.2 & 25.87 & & 44.41 & 72.67 & 89.96 \\
& SPLADE-v2 & 18.4 & 44.1 & 71.3 & 89.6 & 26.40 & & 45.28 & 73.41 & 90.38 \\
& SP\textsubscript{Qwen-backbone} & 18.2 & 43.7 & 72.3 & 91.3 & 25.46 & & 45.12 & 73.46 & 91.79 \\
& SP\textsubscript{Qwen-echoembedding} & 19.7 & 44.0 & 74.2 & 92.4 & 26.72 & & 45.27 & 74.35 & 92.84 \\
& SP\textsubscript{Qwen-bidattention} & 19.5 & 43.8 & 73.5 & 91.8 & 26.41 & & 45.39 & 73.89 & 92.10 \\
\midrule
\multirow{5}{*}[-1pt]{\textit{Ours}}
& PROSPER & 25.3 & 50.7 & 78.1 & 93.9 & 32.85 & & 50.90 & 76.20 & 94.08 \\
& PROSPER\textsubscript{BERT} & 23.6 & 49.2 & 75.8 & 91.5 & 31.83 & & 50.43 & 75.97 & 93.68 \\
& PROSPER\textsubscript{max-pooling} & 22.6 & 49.5 & 78.3 & 93.6 & 30.85 & & 48.45 & 76.33 & 93.75 \\
& PROSPER\textsubscript{echo-emb} & 24.5 & 50.7 & 80.3 & \textbf{94.1} & 32.82 & & 50.18 & \textbf{77.20} & \textbf{94.43} \\
& PROSPER\textsubscript{bid-attention} & 22.3 & 51.4 & 78.7 & 93.6 & 30.72 & & \textbf{51.06} & 76.51 & 93.80 \\
\bottomrule
\end{tabular}
\vspace{-2mm}
\end{table*}

\heading{\emph{Implementation details}}
We use Qwen2.5-3B as the default backbone model with lexical focusing window sizes of $k_q=256$ and $k_d=512$.  All models are trained for five epochs using eight NVIDIA H20 96GB GPUs with a learning rate of $3e^{-5}$, batch size of 64 per device, and AdamW optimizer \cite{loshchilov2019decoupledweightdecayregularization}. The FLOPS regularization parameters $\lambda_q$ and $\lambda_d$ are set to $5e^{-3}$ and $1e^{-3}$, respectively. For all analytical experiments beyond overall performance comparison, we conduct experiments exclusively on the public Multi-CPR E-commerce dataset to facilitate reproducibility and enable direct comparison with future research. More detailed implementation configurations are provided in Appendix~\ref{sec:implement}.

\noindent \textbf{\emph{Evaluation metrics}}
We employ different evaluation metrics tailored to the characteristics of each dataset:
\begin{enumerate*}[label=(\roman*)]
\item For Multi-CPR E-commerce dataset, where each query corresponds to a single relevant item, we use MRR@10 and Hit@k (for k=1, 10, 100, 1000);
\item For Taobao-Internal dataset, where each query may have multiple (1 to 10) relevant items, we use Recall@k (for k=10, 100, 1000).  
\end{enumerate*}
In product search, Recall@1000 measure the system's capability to retrieve relevant items within a larger candidate set, which is crucial for first-stage retrieval.

\subsection{Experimental results}
\heading{Overall performance} 
We conduct a comprehensive comparison of PROSPER against various baselines. As shown in Table \ref{tab:main_results}, the main results in both datasets reveal several key findings:
\begin{enumerate*}[label=(\roman*)]
\item Our proposed PROSPER consistently and significantly outperforms all sparse baselines across both datasets. This demonstrates the superior effectiveness of leveraging LLMs for sparse retrieval in real-world scenarios. It is worth noting that PROSPER\textsubscript{BERT} also significantly outperforms SPLADE-v2, indicating that our proposed LRN is effective not only with Qwen2.5 but also with BERT models. This suggests that our approach effectively mitigates the lexical expansion hallucination problem regardless of the backbone architecture;
\item Compared to dense baselines, PROSPER achieves comparable performance. Advanced dense models like the BGE series achieve superior performance in Hit@1,10,100 and MRR@10 on the public dataset, demonstrating the advantage of dense retrieval in precisely ranking target products at top positions. However, PROSPER shows superior performance on Hit@1000 and Recall@1000 metrics, indicating better coverage of relevant items in the candidate set. This advantage is particularly crucial for first-stage retrieval, where the primary goal is to ensure that relevant items are included for subsequent ranking stages rather than to achieve a perfect initial ranking;
\item The variants further validate PROSPER. The performance gap between PROSPER and PROSPER\textsubscript{BERT} highlights the significant advantage of using Qwen2.5 as the backbone. Comparing PROSPER with PROSPER\textsubscript{max-pooling}, we observe that last-token pooling strategy outperforms the max-pooling approach used in SPLADE, particularly on Hit@1 and MRR@10 metrics. The PROSPER\textsubscript{echo-emb} variant shows that repeating input sequences can better utilize information across the entire sequence, with modest performance gains especially on Hit@1000 and Recall metrics. But the higher computational cost of sequence duplication also requires more caution in actually deploying this approach. Interestingly, when switching to bidirectional attention in PROSPER\textsubscript{bid-attention}, we find no consistent improvement over the causal attention model. We hypothesize that forcibly changing the attention mechanism to bidirectional may disrupt the pre-trained knowledge in the LLM.
\end{enumerate*}

\begin{figure}[t]
\centering
\includegraphics[width=\columnwidth]{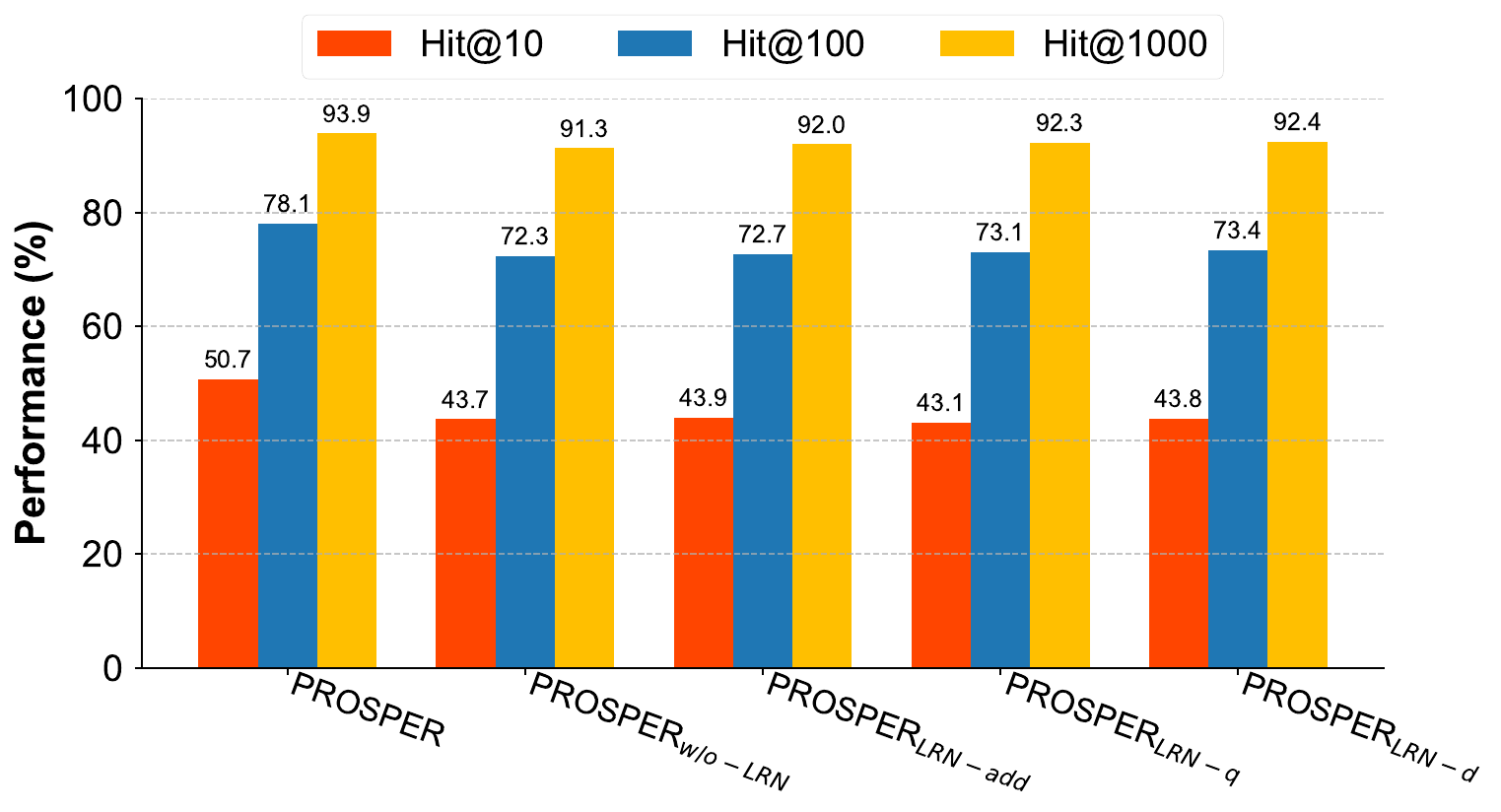}
\caption{Impact of the LRN on retrieval performance.}
\label{fig:lrn_impact}
\vspace{-2mm}
\end{figure}

\heading{Impact of LRN} 
To deeply evaluate the effectiveness of our proposed LRN, we conduct a series of experiments with different variants. As shown in Figure \ref{fig:lrn_impact}, the results clearly demonstrate the critical role of our proposed LRN mechanism in improving retrieval performance. Removing LRN (PROSPER\textsubscript{w/o-LRN}) leads to a dramatic performance drop across all metrics. Interestingly, both PROSPER\textsubscript{LRN-q} and PROSPER\textsubscript{LRN-d} achieve similar performance improvements over the baseline without LRN, indicating that the LRN mechanism is beneficial for both query and item representations. Furthermore, comparing PROSPER with PROSPER\textsubscript{LRN-add} reveals the advantage of our compensatory weighting mechanism over the more intuitive direct addition approach.

\begin{table}[t]
\centering
\caption{Impact of the LFW on retrieval performance(\%).}
\label{tab:lfw_impact}
\setlength{\tabcolsep}{2pt}
\begin{tabular}{@{}l@{\hspace{4pt}}c@{\hspace{4pt}}c@{\hspace{4pt}}c@{\hspace{4pt}}c@{}}
\toprule
\textbf{Variant} & \textbf{Hit@1} & \textbf{Hit@10} & \textbf{Hit@100} & \textbf{Hit@1000} \\
\midrule
$k_d$=1024, $k_q$=512 & 22.3 & 49.2 & 77.7 & 93.5 \\
$k_d$=512, $k_q$=256 & 25.3 & 50.7 & 78.1 & 93.9 \\
$k_d$=256, $k_q$=128 & 23.7 & 50.2 & 77.0 & 93.4 \\
$k_d$=128, $k_q$=64 & 23.5 & 49.1 & 77.2 & 92.9 \\
$k_d$=64, $k_q$=32 & 21.9 & 48.3 & 74.3 & 91.9 \\
PROSPER\textsubscript{LFW-dynamic} & 24.6 & 49.0 & 77.8 & 93.4 \\
PROSPER\textsubscript{w/o-LFW} & 13.3 & 36.2 & 65.6 & 87.2 \\
\bottomrule
\end{tabular}
\end{table}

\heading{Impact of LFW}
To evaluate the effectiveness of LFW, we evaluate PROSPER with different focusing window size configurations. As show in Table \ref{tab:lfw_impact}, we find that:
\begin{enumerate*}[label=(\roman*)]
\item Without LFW (PROSPER\textsubscript{w/o-LFW}), the model's performance drops dramatically across all metrics. This confirms our hypothesis that guiding the model's attention to the top relevant terms during training is essential;
\item Among the different window sizes, the setting $k_q$ = 512, $k_d$ = 256 achieves the best overall performance, striking an optimal balance between focusing on the most relevant terms and maintaining sufficient vocabulary coverage;
\item The dynamical LFW approach PROSPER\textsubscript{LFW-dynamic} ($(256, 128, 64)$ for queries and $(512, 256, 128)$ for items) does not exceed the optimal fixed window configuration. This suggests that a properly chosen fixed window size is sufficient for effective training, and the additional complexity of dynamically adjusting window sizes may not provide significant benefits.
\end{enumerate*}
Moreover, we explore the sparsity evolution of query and item representations during training with and without LFW, and find that LFW enables rapid dimension reduction in early training stages while achieving faster stabilization (see Appendix~\ref{sec:sparsity_analysis}).

\begin{figure}[h]
\centering
\includegraphics[width=\linewidth]{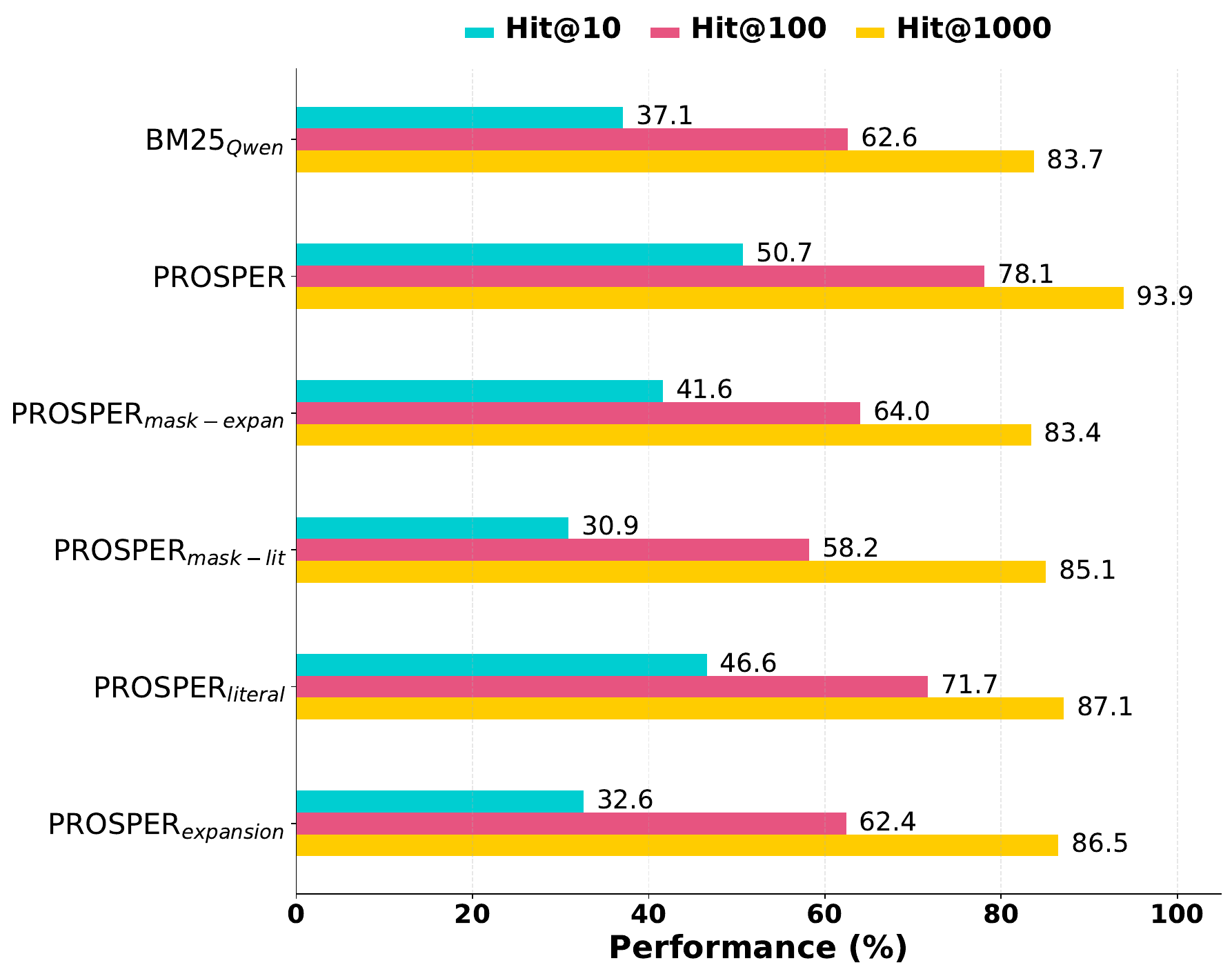}
\caption{Analysis of the impact of literal terms and expansion terms on retrieval performance.}
\label{fig:literal_vs_expansion}
\end{figure}

\heading{Literal terms vs expansion terms}
To analyze the contributions of literal and expansion terms, we compare several variants in Figure~\ref{fig:literal_vs_expansion} and find:
\begin{enumerate*}[label=(\roman*)]
\item The PROSPER\textsubscript{literal} model, trained only in literal terms, significantly outperforms the strong BM25 baseline (e.g., +9.5 points in Hit@10), demonstrating the effectiveness of our model in evaluating the importance of terms even without expansion;
\item Literal terms are superior for precision. This is evident as PROSPER\textsubscript{literal} and PROSPER\textsubscript{mask-expan} significantly outperform their expansion-only counterpart, PROSPER\textsubscript{expansion} and PROSPER\textsubscript{mask-lit}, in metrics such as Hit@10 and Hit@100;
\item Expansion terms are crucial for supplementing recall. When trained in isolation, the recall of PROSPER\textsubscript{expansion} nearly matches that of PROSPER\textsubscript{literal} at Hit@1000. More compellingly, when evaluating the fully trained model by masking components, the expansion-only variant (PROSPER\textsubscript{mask-lit}) even surpasses the literal-only variant (PROSPER\textsubscript{mask-expan}) on the Hit@1000.
\end{enumerate*}

In summary, literal and expansion terms have complementary strengths. We find that using either literal or expansion terms alone can achieve reasonable recall, but they have different advantages. PROSPER masterfully integrates these two aspects, thereby achieving superior overall performance.

\begin{figure}[t]
    \centering
    \includegraphics[width=\linewidth]{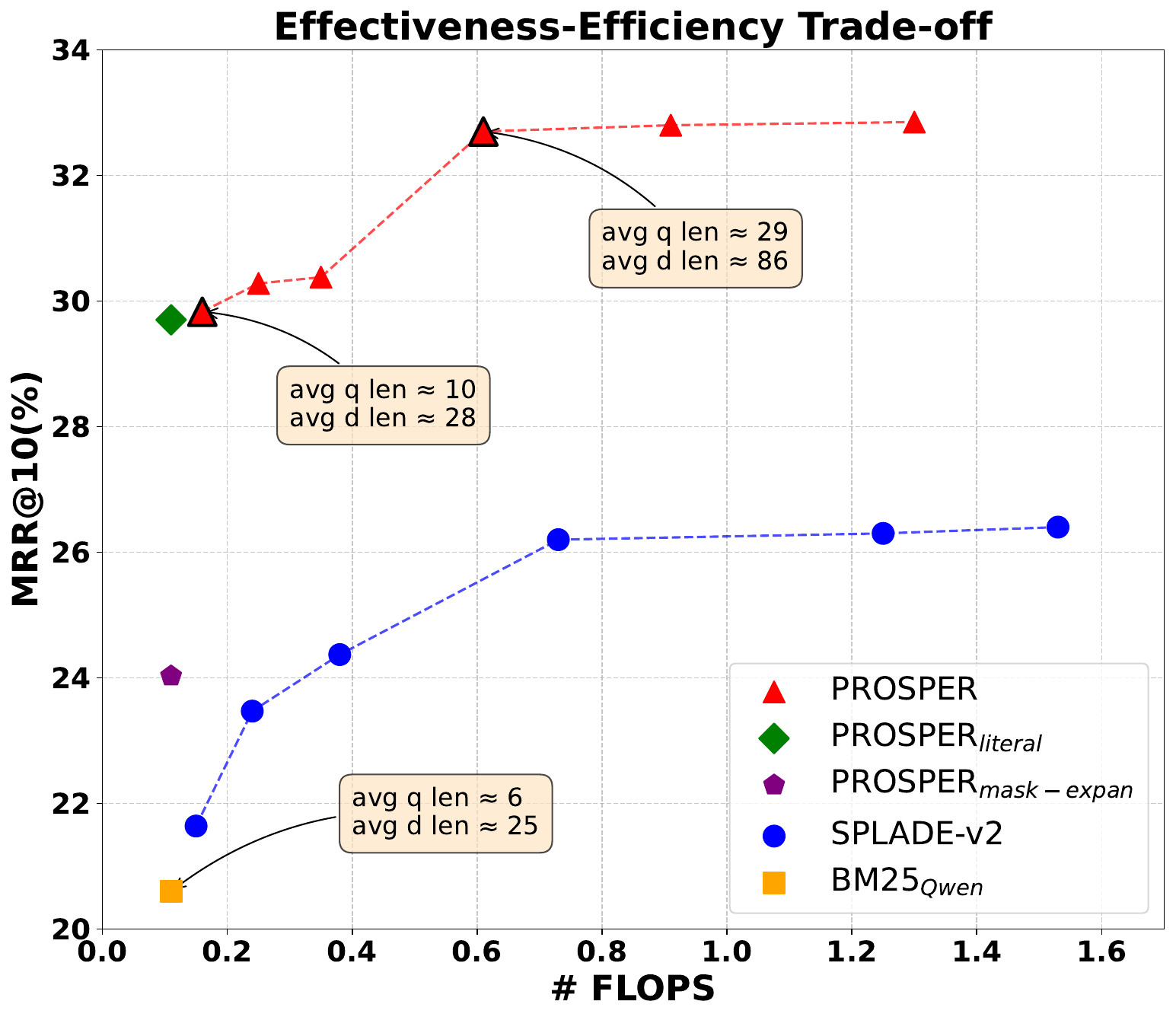}
    \caption{Effectiveness (MRR@10) vs. Efficiency (\# FLOPS) trade-off on Multi-CPR E-commerce. }
    \label{fig:efficiency_tradeoff}
    \vspace{-2mm}
\end{figure}

\heading{Effectiveness-efficiency trade-off}
A critical aspect of industrial search systems is the trade-off between retrieval effectiveness and computational efficiency. We evaluate this trade-off by plotting model performance (MRR@10) against computational cost, measured by \# FLOPS \cite{10.1145/3404835.3463098}. In this context, \# FLOPS represents the average number of overlapping terms between query and item representations. We control \# FLOPS by adjusting the FLOPS regularization strength and the LFW size. As shown in Figure~\ref{fig:efficiency_tradeoff}, we find that:
\begin{enumerate*}[label=(\roman*)]
\item Both PROSPER and SPLADE can achieve strong performance at low computational costs, but PROSPER consistently demonstrates a superior trade-off;
\item PROSPER shows immense potential in balancing efficiency and effectiveness. Compared to the BM25, PROSPER\textsubscript{literal} and PROSPER\textsubscript{mask-expan} achieve a dramatic improvement in retrieval quality at a comparable cost. 
\end{enumerate*}

Furthermore, we investigate model size scaling and find no significant performance gains with larger models (see Appendix~\ref{sec:para_scal}), leading us to select the Qwen2.5-1.5B model as the backbone for subsequent online experiments.

\heading{Offline case study}
To provide concrete insights into how PROSPER optimizes term expansion and weighting, we present detailed offline case studies in the Appendix~\ref{sec:offline_case_study}.
We showcases the optimization results for the two examples mentioned in the section \ref{subsec:Application and Challenges}. Additionally, the appendix includes more comprehensive examples of expansion and weighting results.
\vspace*{-1mm}
\section{Online Experiments}
\label{sec:online_experiment}
In this section, we introduce the deployment of PROSPER in the Taobao search engine and present the corresponding online evaluation results and analysis.

\subsection{System deployment}
\begin{figure}[t]
\centering
\includegraphics[width=\linewidth]{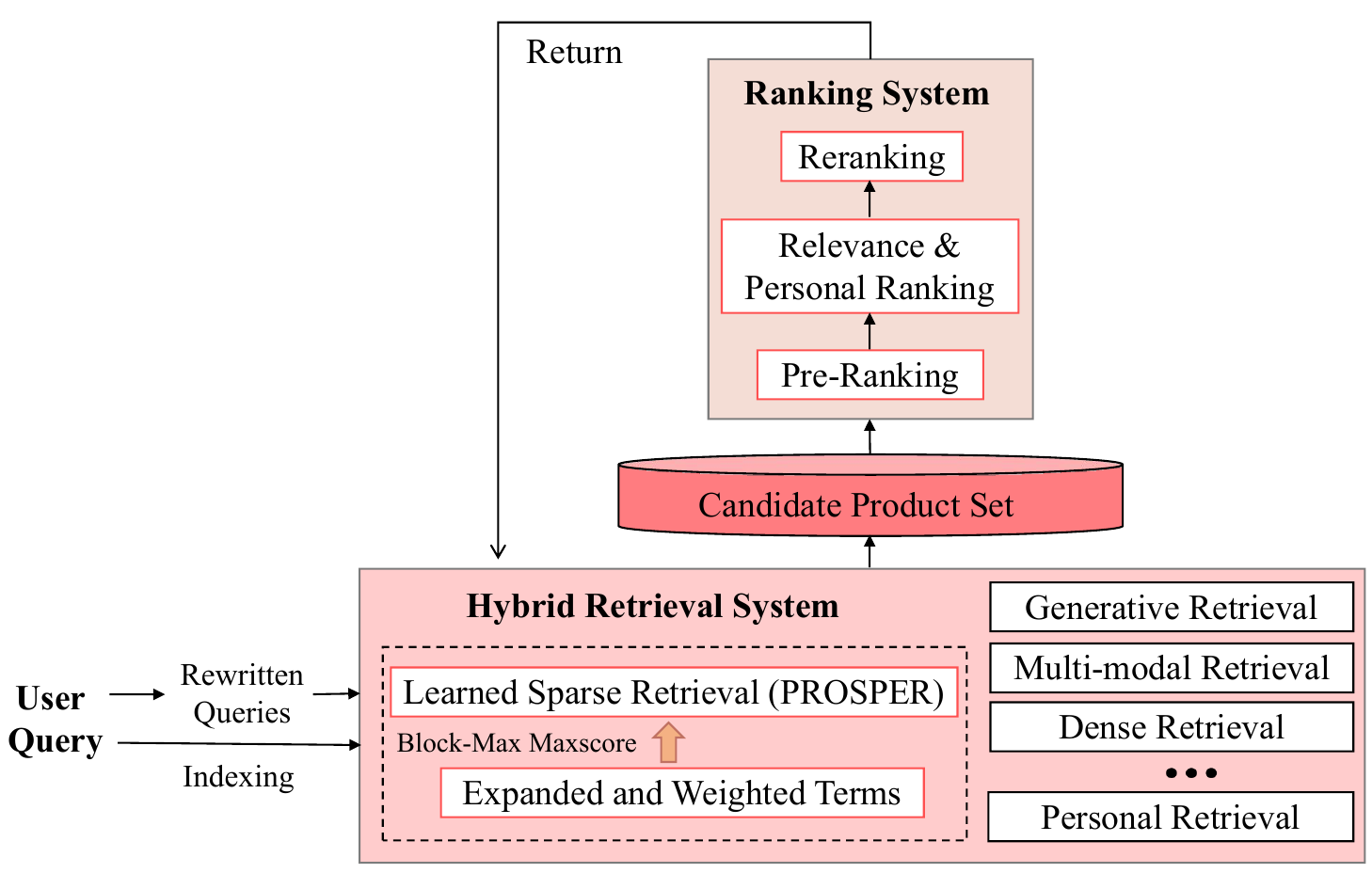}
\caption{Overview of the architecture of the Taobao search
engine with the proposed sparse retrieval model PROSPER.}
\label{fig:system}
\end{figure}

Taobao search engine follows the paradigm of ``index-retrieve-then-rank'' \cite{10.1145/3637528.3671654}. 
As shown in Figure ~\ref{fig:system}, we deploy the proposed sparse retrieval model PROSPER to the retrieval system of the Taobao search engine. 
The first-stage product retrieval adopts a hybrid system combining PROSPER (learned sparse retrieval) with traditional inverted index, multi-modal, generative, dense, and personalized retrieval to comprehensively meet diverse user needs.

When deploying PROSPER for first-stage product retrieval, we first perform offline inference on the product corpus, associating products with corresponding term inverted lists and storing weights to construct the inverted index. During the query process, user queries are input into the model to obtain their represented terms, which are then used to retrieve relevant items through inverted index lookup. This process utilizes the Block-Max Maxscore \cite{10.1145/2433396.2433412,mallia2024fasterlearnedsparseretrieval,10.1145/3637528.3671654} algorithm to optimize efficiency. Finally, we obtain a highly relevant candidate document set for subsequent ranking stages.

\subsection{Online experimental setup}
We select Qwen2.5-1.5B as our backbone model for PROSPER implementation. For training data construction, we sample approximately 330 million real user click records from Taobao search in July 2025 as our training set and train the model for one epoch. For online inference, we perform offline inference on approximately 80 million high-quality products from the product pool and 170 million user queries. To optimize online matching efficiency, we only retain the top-16 weighted terms from the query side for retrieval.

We deploy our inference results in a real online traffic bucket in the Taobao search (approximately 1\% of total traffic), using another online traffic bucket without PROSPER deployment as a baseline for comparison. Both buckets maintain identical configurations except for the presence or absence of PROSPER deployment. All other experimental settings remain consistent with the offline experiments.

For online evaluation, we employ several key metrics: total guided net gross merchandise volume (TG-GMV), direct guided net gross merchandise volume (DG-GMV), net unique visitor (UV), and conversion rate (CVR). Direct guided refers to users making purchases immediately after searching, while total guided encompasses multiple purchase channels such as direct purchases, live streaming purchases, and advertisement-guided purchases.

\subsection{Online experimental results}
\heading{Online A/B tests results}
After conducting a 10-day A/B test, our bucket deployed with PROSPER showed significant improvements in several key metrics compared to the baseline bucket. The detailed results are presented in Table~\ref{tab:online_ab_results}. Beyond these key metrics, other metrics also demonstrated positive trends, and importantly, there was no negative impact on metrics from other stages (such as ranking, advertising, and recommendation), achieving excellent online incremental benefits.
\begin{table}[h]
\centering
\caption{Online A/B test results}
\label{tab:online_ab_results}
\begin{tabular}{ccccc}
\toprule
\textbf{Metric} & \textbf{TG-GMV} & \textbf{DG-GMV} & \textbf{UV} & \textbf{CVR} \\
\midrule
\textbf{Improvement} & +0.64\% & +0.28\% & +0.19\% & +0.22\% \\
\bottomrule
\end{tabular}
\vspace{-2mm}
\end{table}

\heading{Online case study}
Beyond online metrics, we also analyze cases of products recalled by PROSPER in the online environment. Through extensive case analysis, we find that PROSPER demonstrates a substantial number of exclusive recall results, effectively supplementing products that were not recalled by other retrieval methods but were deemed relevant to user needs after evaluation by Taobao's internal query-document relevance analysis model (see Appendix~\ref{sec:online_case_study} for detailed online case studies). This demonstrates that PROSPER can effectively complement recall performance, enabling the search engine to better satisfy user needs.
\section{Conclusion}
\label{sec:conclusion}

In this paper, we explore the application of LLMs for learned sparse retrieval in product search.
Through the LRN and LFW, our proposed PROSPER framework effectively addresses lexical expansion hallucination and training initialization challenges, achieving improvements in both offline and online experimental results. 

\heading{Limitations and future works}
While PROSPER performs sufficiently well, directly leveraging model logits for term expansion and weighting still inevitably introduces some noise terms.  In future work, we plan to explore incorporating CoT \cite{10.5555/3600270.3602070} reasoning to filter   and refine expansion terms. Additionally, we will investigate the application of learned sparse retrieval in ranking stages of product search pipelines, exploring how sparse representations can enhance multi-stage retrieval architectures.

\newpage
\bibliographystyle{ACM-Reference-Format}
\balance
\bibliography{references}


\begin{thebibliography}{55}


\ifx \showCODEN    \undefined \def \showCODEN     #1{\unskip}     \fi
\ifx \showISBNx    \undefined \def \showISBNx     #1{\unskip}     \fi
\ifx \showISBNxiii \undefined \def \showISBNxiii  #1{\unskip}     \fi
\ifx \showISSN     \undefined \def \showISSN      #1{\unskip}     \fi
\ifx \showLCCN     \undefined \def \showLCCN      #1{\unskip}     \fi
\ifx \shownote     \undefined \def \shownote      #1{#1}          \fi
\ifx \showarticletitle \undefined \def \showarticletitle #1{#1}   \fi
\ifx \showURL      \undefined \def \showURL       {\relax}        \fi
\providecommand\bibfield[2]{#2}
\providecommand\bibinfo[2]{#2}
\providecommand\natexlab[1]{#1}
\providecommand\showeprint[2][]{arXiv:#2}

\bibitem[935(2021)]%
        {9357643}
 \bibinfo{year}{2021}\natexlab{}.
\newblock \bibinfo{booktitle}{\emph{35 A Statistical Interpretation of Term Specificity and Its Application in Retrieval (1972)}}.
\newblock \bibinfo{pages}{339--347}.
\newblock


\bibitem[Bai et~al\mbox{.}(2020)]%
        {baiSparTermLearningTermbased2020}
\bibfield{author}{\bibinfo{person}{Yang Bai}, \bibinfo{person}{Xiaoguang Li}, \bibinfo{person}{Gang Wang}, \bibinfo{person}{Chaoliang Zhang}, \bibinfo{person}{Lifeng Shang}, \bibinfo{person}{Jun Xu}, \bibinfo{person}{Zhaowei Wang}, \bibinfo{person}{Fangshan Wang}, {and} \bibinfo{person}{Qun Liu}.} \bibinfo{year}{2020}\natexlab{}.
\newblock \bibinfo{title}{{{SparTerm}}: {{Learning Term-based Sparse Representation}} for {{Fast Text Retrieval}}}.
\newblock
\showeprint[arxiv]{2010.00768}~[cs]
\href{https://doi.org/10.48550/arXiv.2010.00768}{doi:\nolinkurl{10.48550/arXiv.2010.00768}}


\bibitem[Bajaj et~al\mbox{.}(2018)]%
        {bajaj2018msmarcohumangenerated}
\bibfield{author}{\bibinfo{person}{Payal Bajaj}, \bibinfo{person}{Daniel Campos}, \bibinfo{person}{Nick Craswell}, \bibinfo{person}{Li Deng}, \bibinfo{person}{Jianfeng Gao}, \bibinfo{person}{Xiaodong Liu}, \bibinfo{person}{Rangan Majumder}, \bibinfo{person}{Andrew McNamara}, \bibinfo{person}{Bhaskar Mitra}, \bibinfo{person}{Tri Nguyen}, \bibinfo{person}{Mir Rosenberg}, \bibinfo{person}{Xia Song}, \bibinfo{person}{Alina Stoica}, \bibinfo{person}{Saurabh Tiwary}, {and} \bibinfo{person}{Tong Wang}.} \bibinfo{year}{2018}\natexlab{}.
\newblock \bibinfo{title}{MS MARCO: A Human Generated MAchine Reading COmprehension Dataset}.
\newblock
\showeprint[arxiv]{1611.09268}~[cs.CL]
\urldef\tempurl%
\url{https://arxiv.org/abs/1611.09268}
\showURL{%
\tempurl}


\bibitem[BehnamGhader et~al\mbox{.}(2024)]%
        {behnamghaderLLM2VecLargeLanguage2024}
\bibfield{author}{\bibinfo{person}{Parishad BehnamGhader}, \bibinfo{person}{Vaibhav Adlakha}, \bibinfo{person}{Marius Mosbach}, \bibinfo{person}{Dzmitry Bahdanau}, \bibinfo{person}{Nicolas Chapados}, {and} \bibinfo{person}{Siva Reddy}.} \bibinfo{year}{2024}\natexlab{}.
\newblock \bibinfo{title}{{{LLM2Vec}}: {{Large Language Models Are Secretly Powerful Text Encoders}}}.
\newblock
\showeprint[arxiv]{2404.05961}~[cs]
\href{https://doi.org/10.48550/arXiv.2404.05961}{doi:\nolinkurl{10.48550/arXiv.2404.05961}}


\bibitem[Dai and Callan(2019a)]%
        {dai2019contextawaresentencepassagetermimportance}
\bibfield{author}{\bibinfo{person}{Zhuyun Dai} {and} \bibinfo{person}{Jamie Callan}.} \bibinfo{year}{2019}\natexlab{a}.
\newblock \bibinfo{title}{Context-Aware Sentence/Passage Term Importance Estimation For First Stage Retrieval}.
\newblock
\showeprint[arxiv]{1910.10687}~[cs.IR]
\urldef\tempurl%
\url{https://arxiv.org/abs/1910.10687}
\showURL{%
\tempurl}


\bibitem[Dai and Callan(2019b)]%
        {10.1145/3331184.3331303}
\bibfield{author}{\bibinfo{person}{Zhuyun Dai} {and} \bibinfo{person}{Jamie Callan}.} \bibinfo{year}{2019}\natexlab{b}.
\newblock \showarticletitle{Deeper Text Understanding for IR with Contextual Neural Language Modeling}. In \bibinfo{booktitle}{\emph{Proceedings of the 42nd International ACM SIGIR Conference on Research and Development in Information Retrieval}} (Paris, France) \emph{(\bibinfo{series}{SIGIR'19})}. \bibinfo{publisher}{Association for Computing Machinery}, \bibinfo{address}{New York, NY, USA}, \bibinfo{pages}{985–988}.
\newblock
\showISBNx{9781450361729}
\href{https://doi.org/10.1145/3331184.3331303}{doi:\nolinkurl{10.1145/3331184.3331303}}


\bibitem[Dai and Callan(2020a)]%
        {10.1145/3366423.3380258}
\bibfield{author}{\bibinfo{person}{Zhuyun Dai} {and} \bibinfo{person}{Jamie Callan}.} \bibinfo{year}{2020}\natexlab{a}.
\newblock \showarticletitle{Context-Aware Document Term Weighting for Ad-Hoc Search}. In \bibinfo{booktitle}{\emph{Proceedings of The Web Conference 2020}} (Taipei, Taiwan) \emph{(\bibinfo{series}{WWW '20})}. \bibinfo{publisher}{Association for Computing Machinery}, \bibinfo{address}{New York, NY, USA}, \bibinfo{pages}{1897–1907}.
\newblock
\showISBNx{9781450370233}
\href{https://doi.org/10.1145/3366423.3380258}{doi:\nolinkurl{10.1145/3366423.3380258}}


\bibitem[Dai and Callan(2020b)]%
        {10.1145/3397271.3401204}
\bibfield{author}{\bibinfo{person}{Zhuyun Dai} {and} \bibinfo{person}{Jamie Callan}.} \bibinfo{year}{2020}\natexlab{b}.
\newblock \showarticletitle{Context-Aware Term Weighting For First Stage Passage Retrieval}. In \bibinfo{booktitle}{\emph{Proceedings of the 43rd International ACM SIGIR Conference on Research and Development in Information Retrieval}} (Virtual Event, China) \emph{(\bibinfo{series}{SIGIR '20})}. \bibinfo{publisher}{Association for Computing Machinery}, \bibinfo{address}{New York, NY, USA}, \bibinfo{pages}{1533–1536}.
\newblock
\showISBNx{9781450380164}
\href{https://doi.org/10.1145/3397271.3401204}{doi:\nolinkurl{10.1145/3397271.3401204}}


\bibitem[Devlin et~al\mbox{.}(2019)]%
        {devlin-etal-2019-bert}
\bibfield{author}{\bibinfo{person}{Jacob Devlin}, \bibinfo{person}{Ming-Wei Chang}, \bibinfo{person}{Kenton Lee}, {and} \bibinfo{person}{Kristina Toutanova}.} \bibinfo{year}{2019}\natexlab{}.
\newblock \showarticletitle{{BERT}: Pre-training of Deep Bidirectional Transformers for Language Understanding}. In \bibinfo{booktitle}{\emph{Proceedings of the 2019 Conference of the North {A}merican Chapter of the Association for Computational Linguistics: Human Language Technologies, Volume 1 (Long and Short Papers)}}, \bibfield{editor}{\bibinfo{person}{Jill Burstein}, \bibinfo{person}{Christy Doran}, {and} \bibinfo{person}{Thamar Solorio}} (Eds.). \bibinfo{publisher}{Association for Computational Linguistics}, \bibinfo{address}{Minneapolis, Minnesota}, \bibinfo{pages}{4171--4186}.
\newblock
\href{https://doi.org/10.18653/v1/N19-1423}{doi:\nolinkurl{10.18653/v1/N19-1423}}


\bibitem[Dimopoulos et~al\mbox{.}(2013)]%
        {10.1145/2433396.2433412}
\bibfield{author}{\bibinfo{person}{Constantinos Dimopoulos}, \bibinfo{person}{Sergey Nepomnyachiy}, {and} \bibinfo{person}{Torsten Suel}.} \bibinfo{year}{2013}\natexlab{}.
\newblock \showarticletitle{Optimizing top-k document retrieval strategies for block-max indexes}. In \bibinfo{booktitle}{\emph{Proceedings of the Sixth ACM International Conference on Web Search and Data Mining}} (Rome, Italy) \emph{(\bibinfo{series}{WSDM '13})}. \bibinfo{publisher}{Association for Computing Machinery}, \bibinfo{address}{New York, NY, USA}, \bibinfo{pages}{113–122}.
\newblock
\showISBNx{9781450318693}
\href{https://doi.org/10.1145/2433396.2433412}{doi:\nolinkurl{10.1145/2433396.2433412}}


\bibitem[Formal et~al\mbox{.}(2021a)]%
        {formalSPLADEV2Sparse2021}
\bibfield{author}{\bibinfo{person}{Thibault Formal}, \bibinfo{person}{Carlos Lassance}, \bibinfo{person}{Benjamin Piwowarski}, {and} \bibinfo{person}{St{\'e}phane Clinchant}.} \bibinfo{year}{2021}\natexlab{a}.
\newblock \bibinfo{title}{{{SPLADE}} v2: {{Sparse Lexical}} and {{Expansion Model}} for {{Information Retrieval}}}.
\newblock
\showeprint[arxiv]{2109.10086}~[cs]
\href{https://doi.org/10.48550/arXiv.2109.10086}{doi:\nolinkurl{10.48550/arXiv.2109.10086}}


\bibitem[Formal et~al\mbox{.}(2022)]%
        {10.1145/3477495.3531857}
\bibfield{author}{\bibinfo{person}{Thibault Formal}, \bibinfo{person}{Carlos Lassance}, \bibinfo{person}{Benjamin Piwowarski}, {and} \bibinfo{person}{St\'{e}phane Clinchant}.} \bibinfo{year}{2022}\natexlab{}.
\newblock \showarticletitle{From Distillation to Hard Negative Sampling: Making Sparse Neural IR Models More Effective}. In \bibinfo{booktitle}{\emph{Proceedings of the 45th International ACM SIGIR Conference on Research and Development in Information Retrieval}} (Madrid, Spain) \emph{(\bibinfo{series}{SIGIR '22})}. \bibinfo{publisher}{Association for Computing Machinery}, \bibinfo{address}{New York, NY, USA}, \bibinfo{pages}{2353–2359}.
\newblock
\showISBNx{9781450387323}
\href{https://doi.org/10.1145/3477495.3531857}{doi:\nolinkurl{10.1145/3477495.3531857}}


\bibitem[Formal et~al\mbox{.}(2021b)]%
        {10.1145/3404835.3463098}
\bibfield{author}{\bibinfo{person}{Thibault Formal}, \bibinfo{person}{Benjamin Piwowarski}, {and} \bibinfo{person}{St\'{e}phane Clinchant}.} \bibinfo{year}{2021}\natexlab{b}.
\newblock \showarticletitle{SPLADE: Sparse Lexical and Expansion Model for First Stage Ranking}. In \bibinfo{booktitle}{\emph{Proceedings of the 44th International ACM SIGIR Conference on Research and Development in Information Retrieval}} (Virtual Event, Canada) \emph{(\bibinfo{series}{SIGIR '21})}. \bibinfo{publisher}{Association for Computing Machinery}, \bibinfo{address}{New York, NY, USA}, \bibinfo{pages}{2288–2292}.
\newblock
\showISBNx{9781450380379}
\href{https://doi.org/10.1145/3404835.3463098}{doi:\nolinkurl{10.1145/3404835.3463098}}


\bibitem[Gao and Callan(2021)]%
        {gaoCondenserPretrainingArchitecture2021}
\bibfield{author}{\bibinfo{person}{Luyu Gao} {and} \bibinfo{person}{Jamie Callan}.} \bibinfo{year}{2021}\natexlab{}.
\newblock \showarticletitle{Condenser: A {{Pre-training Architecture}} for {{Dense Retrieval}}}. In \bibinfo{booktitle}{\emph{Proceedings of the 2021 {{Conference}} on {{Empirical Methods}} in {{Natural Language Processing}}}}, \bibfield{editor}{\bibinfo{person}{Marie-Francine Moens}, \bibinfo{person}{Xuanjing Huang}, \bibinfo{person}{Lucia Specia}, {and} \bibinfo{person}{Scott Wen-tau Yih}} (Eds.). \bibinfo{publisher}{Association for Computational Linguistics}, \bibinfo{address}{Online and Punta Cana, Dominican Republic}, \bibinfo{pages}{981--993}.
\newblock
\href{https://doi.org/10.18653/v1/2021.emnlp-main.75}{doi:\nolinkurl{10.18653/v1/2021.emnlp-main.75}}


\bibitem[Gao and Callan(2022)]%
        {gaoUnsupervisedCorpusAware2022}
\bibfield{author}{\bibinfo{person}{Luyu Gao} {and} \bibinfo{person}{Jamie Callan}.} \bibinfo{year}{2022}\natexlab{}.
\newblock \showarticletitle{Unsupervised {{Corpus Aware Language Model Pre-training}} for {{Dense Passage Retrieval}}}. In \bibinfo{booktitle}{\emph{Proceedings of the 60th {{Annual Meeting}} of the {{Association}} for {{Computational Linguistics}} ({{Volume}} 1: {{Long Papers}})}}, \bibfield{editor}{\bibinfo{person}{Smaranda Muresan}, \bibinfo{person}{Preslav Nakov}, {and} \bibinfo{person}{Aline Villavicencio}} (Eds.). \bibinfo{publisher}{Association for Computational Linguistics}, \bibinfo{address}{Dublin, Ireland}, \bibinfo{pages}{2843--2853}.
\newblock
\href{https://doi.org/10.18653/v1/2022.acl-long.203}{doi:\nolinkurl{10.18653/v1/2022.acl-long.203}}


\bibitem[Gao et~al\mbox{.}(2022)]%
        {gao2022tevatronefficientflexibletoolkit}
\bibfield{author}{\bibinfo{person}{Luyu Gao}, \bibinfo{person}{Xueguang Ma}, \bibinfo{person}{Jimmy Lin}, {and} \bibinfo{person}{Jamie Callan}.} \bibinfo{year}{2022}\natexlab{}.
\newblock \bibinfo{title}{Tevatron: An Efficient and Flexible Toolkit for Dense Retrieval}.
\newblock
\showeprint[arxiv]{2203.05765}~[cs.IR]
\urldef\tempurl%
\url{https://arxiv.org/abs/2203.05765}
\showURL{%
\tempurl}


\bibitem[Guo et~al\mbox{.}(2022)]%
        {guoSemanticModelsFirststage2022}
\bibfield{author}{\bibinfo{person}{Jiafeng Guo}, \bibinfo{person}{Yinqiong Cai}, \bibinfo{person}{Yixing Fan}, \bibinfo{person}{Fei Sun}, \bibinfo{person}{Ruqing Zhang}, {and} \bibinfo{person}{Xueqi Cheng}.} \bibinfo{year}{2022}\natexlab{}.
\newblock \showarticletitle{Semantic {{Models}} for the {{First-stage Retrieval}}: {{A Comprehensive Review}}}.
\newblock \bibinfo{journal}{\emph{ACM Transactions on Information Systems}} \bibinfo{volume}{40}, \bibinfo{number}{4} (\bibinfo{year}{2022}), \bibinfo{pages}{1--42}.
\newblock
\showISSN{1046-8188, 1558-2868}
\showeprint[arxiv]{2103.04831}~[cs]
\href{https://doi.org/10.1145/3486250}{doi:\nolinkurl{10.1145/3486250}}


\bibitem[Hofst\"{a}tter et~al\mbox{.}(2021)]%
        {10.1145/3404835.3462891}
\bibfield{author}{\bibinfo{person}{Sebastian Hofst\"{a}tter}, \bibinfo{person}{Sheng-Chieh Lin}, \bibinfo{person}{Jheng-Hong Yang}, \bibinfo{person}{Jimmy Lin}, {and} \bibinfo{person}{Allan Hanbury}.} \bibinfo{year}{2021}\natexlab{}.
\newblock \showarticletitle{Efficiently Teaching an Effective Dense Retriever with Balanced Topic Aware Sampling}. In \bibinfo{booktitle}{\emph{Proceedings of the 44th International ACM SIGIR Conference on Research and Development in Information Retrieval}} (Virtual Event, Canada) \emph{(\bibinfo{series}{SIGIR '21})}. \bibinfo{publisher}{Association for Computing Machinery}, \bibinfo{address}{New York, NY, USA}, \bibinfo{pages}{113–122}.
\newblock
\showISBNx{9781450380379}
\href{https://doi.org/10.1145/3404835.3462891}{doi:\nolinkurl{10.1145/3404835.3462891}}


\bibitem[Hofstätter et~al\mbox{.}(2021)]%
        {hofstätter2021improvingefficientneuralranking}
\bibfield{author}{\bibinfo{person}{Sebastian Hofstätter}, \bibinfo{person}{Sophia Althammer}, \bibinfo{person}{Michael Schröder}, \bibinfo{person}{Mete Sertkan}, {and} \bibinfo{person}{Allan Hanbury}.} \bibinfo{year}{2021}\natexlab{}.
\newblock \bibinfo{title}{Improving Efficient Neural Ranking Models with Cross-Architecture Knowledge Distillation}.
\newblock
\showeprint[arxiv]{2010.02666}~[cs.IR]
\urldef\tempurl%
\url{https://arxiv.org/abs/2010.02666}
\showURL{%
\tempurl}


\bibitem[Izacard et~al\mbox{.}(2022)]%
        {izacard2022unsuperviseddenseinformationretrieval}
\bibfield{author}{\bibinfo{person}{Gautier Izacard}, \bibinfo{person}{Mathilde Caron}, \bibinfo{person}{Lucas Hosseini}, \bibinfo{person}{Sebastian Riedel}, \bibinfo{person}{Piotr Bojanowski}, \bibinfo{person}{Armand Joulin}, {and} \bibinfo{person}{Edouard Grave}.} \bibinfo{year}{2022}\natexlab{}.
\newblock \bibinfo{title}{Unsupervised Dense Information Retrieval with Contrastive Learning}.
\newblock
\showeprint[arxiv]{2112.09118}~[cs.IR]
\urldef\tempurl%
\url{https://arxiv.org/abs/2112.09118}
\showURL{%
\tempurl}


\bibitem[Johnson et~al\mbox{.}(2021)]%
        {8733051}
\bibfield{author}{\bibinfo{person}{Jeff Johnson}, \bibinfo{person}{Matthijs Douze}, {and} \bibinfo{person}{Hervé Jégou}.} \bibinfo{year}{2021}\natexlab{}.
\newblock \showarticletitle{Billion-Scale Similarity Search with GPUs}.
\newblock \bibinfo{journal}{\emph{IEEE Transactions on Big Data}} \bibinfo{volume}{7}, \bibinfo{number}{3} (\bibinfo{year}{2021}), \bibinfo{pages}{535--547}.
\newblock
\href{https://doi.org/10.1109/TBDATA.2019.2921572}{doi:\nolinkurl{10.1109/TBDATA.2019.2921572}}


\bibitem[Karpukhin et~al\mbox{.}(2020)]%
        {karpukhin-etal-2020-dense}
\bibfield{author}{\bibinfo{person}{Vladimir Karpukhin}, \bibinfo{person}{Barlas Oguz}, \bibinfo{person}{Sewon Min}, \bibinfo{person}{Patrick Lewis}, \bibinfo{person}{Ledell Wu}, \bibinfo{person}{Sergey Edunov}, \bibinfo{person}{Danqi Chen}, {and} \bibinfo{person}{Wen-tau Yih}.} \bibinfo{year}{2020}\natexlab{}.
\newblock \showarticletitle{Dense Passage Retrieval for Open-Domain Question Answering}. In \bibinfo{booktitle}{\emph{Proceedings of the 2020 Conference on Empirical Methods in Natural Language Processing (EMNLP)}}, \bibfield{editor}{\bibinfo{person}{Bonnie Webber}, \bibinfo{person}{Trevor Cohn}, \bibinfo{person}{Yulan He}, {and} \bibinfo{person}{Yang Liu}} (Eds.). \bibinfo{publisher}{Association for Computational Linguistics}, \bibinfo{address}{Online}, \bibinfo{pages}{6769--6781}.
\newblock
\href{https://doi.org/10.18653/v1/2020.emnlp-main.550}{doi:\nolinkurl{10.18653/v1/2020.emnlp-main.550}}


\bibitem[Khattab and Zaharia(2020)]%
        {10.1145/3397271.3401075}
\bibfield{author}{\bibinfo{person}{Omar Khattab} {and} \bibinfo{person}{Matei Zaharia}.} \bibinfo{year}{2020}\natexlab{}.
\newblock \showarticletitle{ColBERT: Efficient and Effective Passage Search via Contextualized Late Interaction over BERT}. In \bibinfo{booktitle}{\emph{Proceedings of the 43rd International ACM SIGIR Conference on Research and Development in Information Retrieval}} (Virtual Event, China) \emph{(\bibinfo{series}{SIGIR '20})}. \bibinfo{publisher}{Association for Computing Machinery}, \bibinfo{address}{New York, NY, USA}, \bibinfo{pages}{39–48}.
\newblock
\showISBNx{9781450380164}
\href{https://doi.org/10.1145/3397271.3401075}{doi:\nolinkurl{10.1145/3397271.3401075}}


\bibitem[Kong et~al\mbox{.}(2023a)]%
        {10.1145/3539618.3592065}
\bibfield{author}{\bibinfo{person}{Weize Kong}, \bibinfo{person}{Jeffrey~M. Dudek}, \bibinfo{person}{Cheng Li}, \bibinfo{person}{Mingyang Zhang}, {and} \bibinfo{person}{Michael Bendersky}.} \bibinfo{year}{2023}\natexlab{a}.
\newblock \showarticletitle{SparseEmbed: Learning Sparse Lexical Representations with Contextual Embeddings for Retrieval}. In \bibinfo{booktitle}{\emph{Proceedings of the 46th International ACM SIGIR Conference on Research and Development in Information Retrieval}} (Taipei, Taiwan) \emph{(\bibinfo{series}{SIGIR '23})}. \bibinfo{publisher}{Association for Computing Machinery}, \bibinfo{address}{New York, NY, USA}, \bibinfo{pages}{2399–2403}.
\newblock
\showISBNx{9781450394086}
\href{https://doi.org/10.1145/3539618.3592065}{doi:\nolinkurl{10.1145/3539618.3592065}}


\bibitem[Kong et~al\mbox{.}(2023b)]%
        {kongSparseEmbedLearningSparse2023}
\bibfield{author}{\bibinfo{person}{Weize Kong}, \bibinfo{person}{Jeffrey~M. Dudek}, \bibinfo{person}{Cheng Li}, \bibinfo{person}{Mingyang Zhang}, {and} \bibinfo{person}{Michael Bendersky}.} \bibinfo{year}{2023}\natexlab{b}.
\newblock \showarticletitle{{{SparseEmbed}}: {{Learning Sparse Lexical Representations}} with {{Contextual Embeddings}} for {{Retrieval}}}. In \bibinfo{booktitle}{\emph{Proceedings of the 46th {{International ACM SIGIR Conference}} on {{Research}} and {{Development}} in {{Information Retrieval}}}} \emph{(\bibinfo{series}{{{SIGIR}} '23})}. \bibinfo{publisher}{Association for Computing Machinery}, \bibinfo{address}{New York, NY, USA}, \bibinfo{pages}{2399--2403}.
\newblock
\showISBNx{978-1-4503-9408-6}
\href{https://doi.org/10.1145/3539618.3592065}{doi:\nolinkurl{10.1145/3539618.3592065}}


\bibitem[Lee et~al\mbox{.}(2025)]%
        {leeNVEmbedImprovedTechniques2025}
\bibfield{author}{\bibinfo{person}{Chankyu Lee}, \bibinfo{person}{Rajarshi Roy}, \bibinfo{person}{Mengyao Xu}, \bibinfo{person}{Jonathan Raiman}, \bibinfo{person}{Mohammad Shoeybi}, \bibinfo{person}{Bryan Catanzaro}, {and} \bibinfo{person}{Wei Ping}.} \bibinfo{year}{2025}\natexlab{}.
\newblock \bibinfo{title}{{{NV-Embed}}: {{Improved Techniques}} for {{Training LLMs}} as {{Generalist Embedding Models}}}.
\newblock
\showeprint[arxiv]{2405.17428}~[cs]
\href{https://doi.org/10.48550/arXiv.2405.17428}{doi:\nolinkurl{10.48550/arXiv.2405.17428}}


\bibitem[Lei et~al\mbox{.}(2023)]%
        {lei-etal-2023-unsupervised}
\bibfield{author}{\bibinfo{person}{Yibin Lei}, \bibinfo{person}{Liang Ding}, \bibinfo{person}{Yu Cao}, \bibinfo{person}{Changtong Zan}, \bibinfo{person}{Andrew Yates}, {and} \bibinfo{person}{Dacheng Tao}.} \bibinfo{year}{2023}\natexlab{}.
\newblock \showarticletitle{Unsupervised Dense Retrieval with Relevance-Aware Contrastive Pre-Training}. In \bibinfo{booktitle}{\emph{Findings of the Association for Computational Linguistics: ACL 2023}}, \bibfield{editor}{\bibinfo{person}{Anna Rogers}, \bibinfo{person}{Jordan Boyd-Graber}, {and} \bibinfo{person}{Naoaki Okazaki}} (Eds.). \bibinfo{publisher}{Association for Computational Linguistics}, \bibinfo{address}{Toronto, Canada}, \bibinfo{pages}{10932--10940}.
\newblock
\href{https://doi.org/10.18653/v1/2023.findings-acl.695}{doi:\nolinkurl{10.18653/v1/2023.findings-acl.695}}


\bibitem[Li et~al\mbox{.}(2023)]%
        {10.1145/3539618.3591977}
\bibfield{author}{\bibinfo{person}{Minghan Li}, \bibinfo{person}{Sheng-Chieh Lin}, \bibinfo{person}{Xueguang Ma}, {and} \bibinfo{person}{Jimmy Lin}.} \bibinfo{year}{2023}\natexlab{}.
\newblock \showarticletitle{SLIM: Sparsified Late Interaction for Multi-Vector Retrieval with Inverted Indexes}. In \bibinfo{booktitle}{\emph{Proceedings of the 46th International ACM SIGIR Conference on Research and Development in Information Retrieval}} (Taipei, Taiwan) \emph{(\bibinfo{series}{SIGIR '23})}. \bibinfo{publisher}{Association for Computing Machinery}, \bibinfo{address}{New York, NY, USA}, \bibinfo{pages}{1954–1959}.
\newblock
\showISBNx{9781450394086}
\href{https://doi.org/10.1145/3539618.3591977}{doi:\nolinkurl{10.1145/3539618.3591977}}


\bibitem[Li et~al\mbox{.}(2024)]%
        {10.1145/3637528.3671654}
\bibfield{author}{\bibinfo{person}{Sen Li}, \bibinfo{person}{Fuyu Lv}, \bibinfo{person}{Ruqing Zhang}, \bibinfo{person}{Dan Ou}, \bibinfo{person}{Zhixuan Zhang}, {and} \bibinfo{person}{Maarten de Rijke}.} \bibinfo{year}{2024}\natexlab{}.
\newblock \showarticletitle{Text Matching Indexers in Taobao Search}. In \bibinfo{booktitle}{\emph{Proceedings of the 30th ACM SIGKDD Conference on Knowledge Discovery and Data Mining}} (Barcelona, Spain) \emph{(\bibinfo{series}{KDD '24})}. \bibinfo{publisher}{Association for Computing Machinery}, \bibinfo{address}{New York, NY, USA}, \bibinfo{pages}{5339–5350}.
\newblock
\showISBNx{9798400704901}
\href{https://doi.org/10.1145/3637528.3671654}{doi:\nolinkurl{10.1145/3637528.3671654}}


\bibitem[Lin et~al\mbox{.}(2021)]%
        {lin-etal-2021-batch}
\bibfield{author}{\bibinfo{person}{Sheng-Chieh Lin}, \bibinfo{person}{Jheng-Hong Yang}, {and} \bibinfo{person}{Jimmy Lin}.} \bibinfo{year}{2021}\natexlab{}.
\newblock \showarticletitle{In-Batch Negatives for Knowledge Distillation with Tightly-Coupled Teachers for Dense Retrieval}. In \bibinfo{booktitle}{\emph{Proceedings of the 6th Workshop on Representation Learning for NLP (RepL4NLP-2021)}}, \bibfield{editor}{\bibinfo{person}{Anna Rogers}, \bibinfo{person}{Iacer Calixto}, \bibinfo{person}{Ivan Vuli{\'c}}, \bibinfo{person}{Naomi Saphra}, \bibinfo{person}{Nora Kassner}, \bibinfo{person}{Oana-Maria Camburu}, \bibinfo{person}{Trapit Bansal}, {and} \bibinfo{person}{Vered Shwartz}} (Eds.). \bibinfo{publisher}{Association for Computational Linguistics}, \bibinfo{address}{Online}, \bibinfo{pages}{163--173}.
\newblock
\href{https://doi.org/10.18653/v1/2021.repl4nlp-1.17}{doi:\nolinkurl{10.18653/v1/2021.repl4nlp-1.17}}


\bibitem[Liu et~al\mbox{.}(2024)]%
        {liuLlama2VecUnsupervisedAdaptation2024}
\bibfield{author}{\bibinfo{person}{Zheng Liu}, \bibinfo{person}{Chaofan Li}, \bibinfo{person}{Shitao Xiao}, \bibinfo{person}{Yingxia Shao}, {and} \bibinfo{person}{Defu Lian}.} \bibinfo{year}{2024}\natexlab{}.
\newblock \showarticletitle{{{Llama2Vec}}: {{Unsupervised Adaptation}} of {{Large Language Models}} for {{Dense Retrieval}}}. In \bibinfo{booktitle}{\emph{Proceedings of the 62nd {{Annual Meeting}} of the {{Association}} for {{Computational Linguistics}} ({{Volume}} 1: {{Long Papers}})}} (Bangkok, Thailand, 2024-08), \bibfield{editor}{\bibinfo{person}{Lun-Wei Ku}, \bibinfo{person}{Andre Martins}, {and} \bibinfo{person}{Vivek Srikumar}} (Eds.). \bibinfo{publisher}{Association for Computational Linguistics}, \bibinfo{pages}{3490--3500}.
\newblock
\href{https://doi.org/10.18653/v1/2024.acl-long.191}{doi:\nolinkurl{10.18653/v1/2024.acl-long.191}}


\bibitem[Long et~al\mbox{.}(2022)]%
        {10.1145/3477495.3531736}
\bibfield{author}{\bibinfo{person}{Dingkun Long}, \bibinfo{person}{Qiong Gao}, \bibinfo{person}{Kuan Zou}, \bibinfo{person}{Guangwei Xu}, \bibinfo{person}{Pengjun Xie}, \bibinfo{person}{Ruijie Guo}, \bibinfo{person}{Jian Xu}, \bibinfo{person}{Guanjun Jiang}, \bibinfo{person}{Luxi Xing}, {and} \bibinfo{person}{Ping Yang}.} \bibinfo{year}{2022}\natexlab{}.
\newblock \showarticletitle{Multi-CPR: A Multi Domain Chinese Dataset for Passage Retrieval}. In \bibinfo{booktitle}{\emph{Proceedings of the 45th International ACM SIGIR Conference on Research and Development in Information Retrieval}} (Madrid, Spain) \emph{(\bibinfo{series}{SIGIR '22})}. \bibinfo{publisher}{Association for Computing Machinery}, \bibinfo{address}{New York, NY, USA}, \bibinfo{pages}{3046–3056}.
\newblock
\showISBNx{9781450387323}
\href{https://doi.org/10.1145/3477495.3531736}{doi:\nolinkurl{10.1145/3477495.3531736}}


\bibitem[Loshchilov and Hutter(2019)]%
        {loshchilov2019decoupledweightdecayregularization}
\bibfield{author}{\bibinfo{person}{Ilya Loshchilov} {and} \bibinfo{person}{Frank Hutter}.} \bibinfo{year}{2019}\natexlab{}.
\newblock \bibinfo{title}{Decoupled Weight Decay Regularization}.
\newblock
\showeprint[arxiv]{1711.05101}~[cs.LG]
\urldef\tempurl%
\url{https://arxiv.org/abs/1711.05101}
\showURL{%
\tempurl}


\bibitem[Ma et~al\mbox{.}(2024)]%
        {10.1145/3626772.3657951}
\bibfield{author}{\bibinfo{person}{Xueguang Ma}, \bibinfo{person}{Liang Wang}, \bibinfo{person}{Nan Yang}, \bibinfo{person}{Furu Wei}, {and} \bibinfo{person}{Jimmy Lin}.} \bibinfo{year}{2024}\natexlab{}.
\newblock \showarticletitle{Fine-Tuning LLaMA for Multi-Stage Text Retrieval}. In \bibinfo{booktitle}{\emph{Proceedings of the 47th International ACM SIGIR Conference on Research and Development in Information Retrieval}} (Washington DC, USA) \emph{(\bibinfo{series}{SIGIR '24})}. \bibinfo{publisher}{Association for Computing Machinery}, \bibinfo{address}{New York, NY, USA}, \bibinfo{pages}{2421–2425}.
\newblock
\showISBNx{9798400704314}
\href{https://doi.org/10.1145/3626772.3657951}{doi:\nolinkurl{10.1145/3626772.3657951}}


\bibitem[MacAvaney et~al\mbox{.}(2020)]%
        {10.1145/3397271.3401262}
\bibfield{author}{\bibinfo{person}{Sean MacAvaney}, \bibinfo{person}{Franco~Maria Nardini}, \bibinfo{person}{Raffaele Perego}, \bibinfo{person}{Nicola Tonellotto}, \bibinfo{person}{Nazli Goharian}, {and} \bibinfo{person}{Ophir Frieder}.} \bibinfo{year}{2020}\natexlab{}.
\newblock \showarticletitle{Expansion via Prediction of Importance with Contextualization}. In \bibinfo{booktitle}{\emph{Proceedings of the 43rd International ACM SIGIR Conference on Research and Development in Information Retrieval}} (Virtual Event, China) \emph{(\bibinfo{series}{SIGIR '20})}. \bibinfo{publisher}{Association for Computing Machinery}, \bibinfo{address}{New York, NY, USA}, \bibinfo{pages}{1573–1576}.
\newblock
\showISBNx{9781450380164}
\href{https://doi.org/10.1145/3397271.3401262}{doi:\nolinkurl{10.1145/3397271.3401262}}


\bibitem[Mallia et~al\mbox{.}(2024)]%
        {mallia2024fasterlearnedsparseretrieval}
\bibfield{author}{\bibinfo{person}{Antonio Mallia}, \bibinfo{person}{Torten Suel}, {and} \bibinfo{person}{Nicola Tonellotto}.} \bibinfo{year}{2024}\natexlab{}.
\newblock \bibinfo{title}{Faster Learned Sparse Retrieval with Block-Max Pruning}.
\newblock
\showeprint[arxiv]{2405.01117}~[cs.IR]
\urldef\tempurl%
\url{https://arxiv.org/abs/2405.01117}
\showURL{%
\tempurl}


\bibitem[Man et~al\mbox{.}(2024)]%
        {man-etal-2024-ullme}
\bibfield{author}{\bibinfo{person}{Hieu Man}, \bibinfo{person}{Nghia~Trung Ngo}, \bibinfo{person}{Franck Dernoncourt}, {and} \bibinfo{person}{Thien~Huu Nguyen}.} \bibinfo{year}{2024}\natexlab{}.
\newblock \showarticletitle{{ULLME}: A Unified Framework for Large Language Model Embeddings with Generation-Augmented Learning}. In \bibinfo{booktitle}{\emph{Proceedings of the 2024 Conference on Empirical Methods in Natural Language Processing: System Demonstrations}}, \bibfield{editor}{\bibinfo{person}{Delia~Irazu Hernandez~Farias}, \bibinfo{person}{Tom Hope}, {and} \bibinfo{person}{Manling Li}} (Eds.). \bibinfo{publisher}{Association for Computational Linguistics}, \bibinfo{address}{Miami, Florida, USA}, \bibinfo{pages}{230--239}.
\newblock
\href{https://doi.org/10.18653/v1/2024.emnlp-demo.24}{doi:\nolinkurl{10.18653/v1/2024.emnlp-demo.24}}


\bibitem[Minaee et~al\mbox{.}(2025)]%
        {minaee2025largelanguagemodelssurvey}
\bibfield{author}{\bibinfo{person}{Shervin Minaee}, \bibinfo{person}{Tomas Mikolov}, \bibinfo{person}{Narjes Nikzad}, \bibinfo{person}{Meysam Chenaghlu}, \bibinfo{person}{Richard Socher}, \bibinfo{person}{Xavier Amatriain}, {and} \bibinfo{person}{Jianfeng Gao}.} \bibinfo{year}{2025}\natexlab{}.
\newblock \bibinfo{title}{Large Language Models: A Survey}.
\newblock
\showeprint[arxiv]{2402.06196}~[cs.CL]
\urldef\tempurl%
\url{https://arxiv.org/abs/2402.06196}
\showURL{%
\tempurl}


\bibitem[Nogueira et~al\mbox{.}(2020)]%
        {nogueiraDocumentRankingPretrained2020}
\bibfield{author}{\bibinfo{person}{Rodrigo Nogueira}, \bibinfo{person}{Zhiying Jiang}, \bibinfo{person}{Ronak Pradeep}, {and} \bibinfo{person}{Jimmy Lin}.} \bibinfo{year}{2020}\natexlab{}.
\newblock \showarticletitle{Document {{Ranking}} with a {{Pretrained Sequence-to-Sequence Model}}}. In \bibinfo{booktitle}{\emph{Findings of the {{Association}} for {{Computational Linguistics}}: {{EMNLP}} 2020}}, \bibfield{editor}{\bibinfo{person}{Trevor Cohn}, \bibinfo{person}{Yulan He}, {and} \bibinfo{person}{Yang Liu}} (Eds.). \bibinfo{publisher}{Association for Computational Linguistics}, \bibinfo{address}{Online}, \bibinfo{pages}{708--718}.
\newblock
\href{https://doi.org/10.18653/v1/2020.findings-emnlp.63}{doi:\nolinkurl{10.18653/v1/2020.findings-emnlp.63}}


\bibitem[Nogueira et~al\mbox{.}(2019)]%
        {nogueira2019documentexpansionqueryprediction}
\bibfield{author}{\bibinfo{person}{Rodrigo Nogueira}, \bibinfo{person}{Wei Yang}, \bibinfo{person}{Jimmy Lin}, {and} \bibinfo{person}{Kyunghyun Cho}.} \bibinfo{year}{2019}\natexlab{}.
\newblock \bibinfo{title}{Document Expansion by Query Prediction}.
\newblock
\showeprint[arxiv]{1904.08375}~[cs.IR]
\urldef\tempurl%
\url{https://arxiv.org/abs/1904.08375}
\showURL{%
\tempurl}


\bibitem[Paria et~al\mbox{.}(2020)]%
        {paria2020minimizingflopslearnefficient}
\bibfield{author}{\bibinfo{person}{Biswajit Paria}, \bibinfo{person}{Chih-Kuan Yeh}, \bibinfo{person}{Ian E.~H. Yen}, \bibinfo{person}{Ning Xu}, \bibinfo{person}{Pradeep Ravikumar}, {and} \bibinfo{person}{Barnabás Póczos}.} \bibinfo{year}{2020}\natexlab{}.
\newblock \bibinfo{title}{Minimizing FLOPs to Learn Efficient Sparse Representations}.
\newblock
\showeprint[arxiv]{2004.05665}~[cs.LG]
\urldef\tempurl%
\url{https://arxiv.org/abs/2004.05665}
\showURL{%
\tempurl}


\bibitem[Qu et~al\mbox{.}(2021)]%
        {qu-etal-2021-rocketqa}
\bibfield{author}{\bibinfo{person}{Yingqi Qu}, \bibinfo{person}{Yuchen Ding}, \bibinfo{person}{Jing Liu}, \bibinfo{person}{Kai Liu}, \bibinfo{person}{Ruiyang Ren}, \bibinfo{person}{Wayne~Xin Zhao}, \bibinfo{person}{Daxiang Dong}, \bibinfo{person}{Hua Wu}, {and} \bibinfo{person}{Haifeng Wang}.} \bibinfo{year}{2021}\natexlab{}.
\newblock \showarticletitle{{R}ocket{QA}: An Optimized Training Approach to Dense Passage Retrieval for Open-Domain Question Answering}. In \bibinfo{booktitle}{\emph{Proceedings of the 2021 Conference of the North American Chapter of the Association for Computational Linguistics: Human Language Technologies}}, \bibfield{editor}{\bibinfo{person}{Kristina Toutanova}, \bibinfo{person}{Anna Rumshisky}, \bibinfo{person}{Luke Zettlemoyer}, \bibinfo{person}{Dilek Hakkani-Tur}, \bibinfo{person}{Iz~Beltagy}, \bibinfo{person}{Steven Bethard}, \bibinfo{person}{Ryan Cotterell}, \bibinfo{person}{Tanmoy Chakraborty}, {and} \bibinfo{person}{Yichao Zhou}} (Eds.). \bibinfo{publisher}{Association for Computational Linguistics}, \bibinfo{address}{Online}, \bibinfo{pages}{5835--5847}.
\newblock
\href{https://doi.org/10.18653/v1/2021.naacl-main.466}{doi:\nolinkurl{10.18653/v1/2021.naacl-main.466}}


\bibitem[Qwen et~al\mbox{.}(2025)]%
        {qwen2025qwen25technicalreport}
\bibfield{author}{\bibinfo{person}{Qwen}, \bibinfo{person}{:}, \bibinfo{person}{An Yang}, \bibinfo{person}{Baosong Yang}, \bibinfo{person}{Beichen Zhang}, \bibinfo{person}{Binyuan Hui}, \bibinfo{person}{Bo Zheng}, \bibinfo{person}{Bowen Yu}, \bibinfo{person}{Chengyuan Li}, \bibinfo{person}{Dayiheng Liu}, \bibinfo{person}{Fei Huang}, \bibinfo{person}{Haoran Wei}, \bibinfo{person}{Huan Lin}, \bibinfo{person}{Jian Yang}, \bibinfo{person}{Jianhong Tu}, \bibinfo{person}{Jianwei Zhang}, \bibinfo{person}{Jianxin Yang}, \bibinfo{person}{Jiaxi Yang}, \bibinfo{person}{Jingren Zhou}, \bibinfo{person}{Junyang Lin}, \bibinfo{person}{Kai Dang}, \bibinfo{person}{Keming Lu}, \bibinfo{person}{Keqin Bao}, \bibinfo{person}{Kexin Yang}, \bibinfo{person}{Le Yu}, \bibinfo{person}{Mei Li}, \bibinfo{person}{Mingfeng Xue}, \bibinfo{person}{Pei Zhang}, \bibinfo{person}{Qin Zhu}, \bibinfo{person}{Rui Men}, \bibinfo{person}{Runji Lin}, \bibinfo{person}{Tianhao Li}, \bibinfo{person}{Tianyi Tang}, \bibinfo{person}{Tingyu Xia},
  \bibinfo{person}{Xingzhang Ren}, \bibinfo{person}{Xuancheng Ren}, \bibinfo{person}{Yang Fan}, \bibinfo{person}{Yang Su}, \bibinfo{person}{Yichang Zhang}, \bibinfo{person}{Yu Wan}, \bibinfo{person}{Yuqiong Liu}, \bibinfo{person}{Zeyu Cui}, \bibinfo{person}{Zhenru Zhang}, {and} \bibinfo{person}{Zihan Qiu}.} \bibinfo{year}{2025}\natexlab{}.
\newblock \bibinfo{title}{Qwen2.5 Technical Report}.
\newblock
\showeprint[arxiv]{2412.15115}~[cs.CL]
\urldef\tempurl%
\url{https://arxiv.org/abs/2412.15115}
\showURL{%
\tempurl}


\bibitem[Robertson and Zaragoza(2009)]%
        {10.1561/1500000019}
\bibfield{author}{\bibinfo{person}{Stephen Robertson} {and} \bibinfo{person}{Hugo Zaragoza}.} \bibinfo{year}{2009}\natexlab{}.
\newblock \showarticletitle{The Probabilistic Relevance Framework: BM25 and Beyond}.
\newblock \bibinfo{journal}{\emph{Found. Trends Inf. Retr.}} \bibinfo{volume}{3}, \bibinfo{number}{4} (\bibinfo{date}{April} \bibinfo{year}{2009}), \bibinfo{pages}{333–389}.
\newblock
\showISSN{1554-0669}
\href{https://doi.org/10.1561/1500000019}{doi:\nolinkurl{10.1561/1500000019}}


\bibitem[Springer et~al\mbox{.}(2024)]%
        {springerRepetitionImprovesLanguage2024}
\bibfield{author}{\bibinfo{person}{Jacob~Mitchell Springer}, \bibinfo{person}{Suhas Kotha}, \bibinfo{person}{Daniel Fried}, \bibinfo{person}{Graham Neubig}, {and} \bibinfo{person}{Aditi Raghunathan}.} \bibinfo{year}{2024}\natexlab{}.
\newblock \bibinfo{title}{Repetition {{Improves Language Model Embeddings}}}.
\newblock
\showeprint[arxiv]{2402.15449}~[cs]
\href{https://doi.org/10.48550/arXiv.2402.15449}{doi:\nolinkurl{10.48550/arXiv.2402.15449}}


\bibitem[Tao et~al\mbox{.}(2024)]%
        {taoLLMsAreAlso2024}
\bibfield{author}{\bibinfo{person}{Chongyang Tao}, \bibinfo{person}{Tao Shen}, \bibinfo{person}{Shen Gao}, \bibinfo{person}{Junshuo Zhang}, \bibinfo{person}{Zhen Li}, \bibinfo{person}{Zhengwei Tao}, {and} \bibinfo{person}{Shuai Ma}.} \bibinfo{year}{2024}\natexlab{}.
\newblock \bibinfo{title}{{{LLMs}} Are {{Also Effective Embedding Models}}: {{An In-depth Overview}}}.
\newblock
\showeprint[arxiv]{2412.12591}~[cs]
\href{https://doi.org/10.48550/arXiv.2412.12591}{doi:\nolinkurl{10.48550/arXiv.2412.12591}}


\bibitem[van~den Oord et~al\mbox{.}(2019)]%
        {oord2019representationlearningcontrastivepredictive}
\bibfield{author}{\bibinfo{person}{Aaron van~den Oord}, \bibinfo{person}{Yazhe Li}, {and} \bibinfo{person}{Oriol Vinyals}.} \bibinfo{year}{2019}\natexlab{}.
\newblock \bibinfo{title}{Representation Learning with Contrastive Predictive Coding}.
\newblock
\showeprint[arxiv]{1807.03748}~[cs.LG]
\urldef\tempurl%
\url{https://arxiv.org/abs/1807.03748}
\showURL{%
\tempurl}


\bibitem[Wang et~al\mbox{.}(2024)]%
        {wang-etal-2024-improving-text}
\bibfield{author}{\bibinfo{person}{Liang Wang}, \bibinfo{person}{Nan Yang}, \bibinfo{person}{Xiaolong Huang}, \bibinfo{person}{Linjun Yang}, \bibinfo{person}{Rangan Majumder}, {and} \bibinfo{person}{Furu Wei}.} \bibinfo{year}{2024}\natexlab{}.
\newblock \showarticletitle{Improving Text Embeddings with Large Language Models}. In \bibinfo{booktitle}{\emph{Proceedings of the 62nd Annual Meeting of the Association for Computational Linguistics (Volume 1: Long Papers)}}, \bibfield{editor}{\bibinfo{person}{Lun-Wei Ku}, \bibinfo{person}{Andre Martins}, {and} \bibinfo{person}{Vivek Srikumar}} (Eds.). \bibinfo{publisher}{Association for Computational Linguistics}, \bibinfo{address}{Bangkok, Thailand}, \bibinfo{pages}{11897--11916}.
\newblock
\href{https://doi.org/10.18653/v1/2024.acl-long.642}{doi:\nolinkurl{10.18653/v1/2024.acl-long.642}}


\bibitem[Wei et~al\mbox{.}(2022)]%
        {10.5555/3600270.3602070}
\bibfield{author}{\bibinfo{person}{Jason Wei}, \bibinfo{person}{Xuezhi Wang}, \bibinfo{person}{Dale Schuurmans}, \bibinfo{person}{Maarten Bosma}, \bibinfo{person}{Brian Ichter}, \bibinfo{person}{Fei Xia}, \bibinfo{person}{Ed~H. Chi}, \bibinfo{person}{Quoc~V. Le}, {and} \bibinfo{person}{Denny Zhou}.} \bibinfo{year}{2022}\natexlab{}.
\newblock \showarticletitle{Chain-of-thought prompting elicits reasoning in large language models}. In \bibinfo{booktitle}{\emph{Proceedings of the 36th International Conference on Neural Information Processing Systems}} (New Orleans, LA, USA) \emph{(\bibinfo{series}{NIPS '22})}. \bibinfo{publisher}{Curran Associates Inc.}, \bibinfo{address}{Red Hook, NY, USA}, Article \bibinfo{articleno}{1800}, \bibinfo{numpages}{14}~pages.
\newblock
\showISBNx{9781713871088}


\bibitem[Xiao et~al\mbox{.}(2023)]%
        {bge_embedding}
\bibfield{author}{\bibinfo{person}{Shitao Xiao}, \bibinfo{person}{Zheng Liu}, \bibinfo{person}{Peitian Zhang}, {and} \bibinfo{person}{Niklas Muennighoff}.} \bibinfo{year}{2023}\natexlab{}.
\newblock \bibinfo{title}{C-Pack: Packaged Resources To Advance General Chinese Embedding}.
\newblock
\showeprint[arxiv]{2309.07597}~[cs.CL]


\bibitem[Xiong et~al\mbox{.}(2020)]%
        {xiong2020approximatenearestneighbornegative}
\bibfield{author}{\bibinfo{person}{Lee Xiong}, \bibinfo{person}{Chenyan Xiong}, \bibinfo{person}{Ye Li}, \bibinfo{person}{Kwok-Fung Tang}, \bibinfo{person}{Jialin Liu}, \bibinfo{person}{Paul Bennett}, \bibinfo{person}{Junaid Ahmed}, {and} \bibinfo{person}{Arnold Overwijk}.} \bibinfo{year}{2020}\natexlab{}.
\newblock \bibinfo{title}{Approximate Nearest Neighbor Negative Contrastive Learning for Dense Text Retrieval}.
\newblock
\showeprint[arxiv]{2007.00808}~[cs.IR]
\urldef\tempurl%
\url{https://arxiv.org/abs/2007.00808}
\showURL{%
\tempurl}


\bibitem[Xu et~al\mbox{.}(2025)]%
        {xuCSPLADELearnedSparse2025}
\bibfield{author}{\bibinfo{person}{Zhichao Xu}, \bibinfo{person}{Aosong Feng}, \bibinfo{person}{Yijun Tian}, \bibinfo{person}{Haibo Ding}, {and} \bibinfo{person}{Lin~Lee Cheong}.} \bibinfo{year}{2025}\natexlab{}.
\newblock \bibinfo{title}{{{CSPLADE}}: {{Learned Sparse Retrieval}} with {{Causal Language Models}}}.
\newblock
\showeprint[arxiv]{2504.10816}~[cs]
\href{https://doi.org/10.48550/arXiv.2504.10816}{doi:\nolinkurl{10.48550/arXiv.2504.10816}}


\bibitem[Zamani et~al\mbox{.}(2018)]%
        {10.1145/3269206.3271800}
\bibfield{author}{\bibinfo{person}{Hamed Zamani}, \bibinfo{person}{Mostafa Dehghani}, \bibinfo{person}{W.~Bruce Croft}, \bibinfo{person}{Erik Learned-Miller}, {and} \bibinfo{person}{Jaap Kamps}.} \bibinfo{year}{2018}\natexlab{}.
\newblock \showarticletitle{From Neural Re-Ranking to Neural Ranking: Learning a Sparse Representation for Inverted Indexing}. In \bibinfo{booktitle}{\emph{Proceedings of the 27th ACM International Conference on Information and Knowledge Management}} (Torino, Italy) \emph{(\bibinfo{series}{CIKM '18})}. \bibinfo{publisher}{Association for Computing Machinery}, \bibinfo{address}{New York, NY, USA}, \bibinfo{pages}{497–506}.
\newblock
\showISBNx{9781450360142}
\href{https://doi.org/10.1145/3269206.3271800}{doi:\nolinkurl{10.1145/3269206.3271800}}


\bibitem[Zeng et~al\mbox{.}(2025)]%
        {zengScalingSparseDense2025}
\bibfield{author}{\bibinfo{person}{Hansi Zeng}, \bibinfo{person}{Julian Killingback}, {and} \bibinfo{person}{Hamed Zamani}.} \bibinfo{year}{2025}\natexlab{}.
\newblock \bibinfo{title}{Scaling {{Sparse}} and {{Dense Retrieval}} in {{Decoder-Only LLMs}}}.
\newblock
\showeprint[arxiv]{2502.15526}~[cs]
\href{https://doi.org/10.48550/arXiv.2502.15526}{doi:\nolinkurl{10.48550/arXiv.2502.15526}}


\bibitem[Zhao et~al\mbox{.}(2021)]%
        {zhao-etal-2021-sparta}
\bibfield{author}{\bibinfo{person}{Tiancheng Zhao}, \bibinfo{person}{Xiaopeng Lu}, {and} \bibinfo{person}{Kyusong Lee}.} \bibinfo{year}{2021}\natexlab{}.
\newblock \showarticletitle{{SPARTA}: Efficient Open-Domain Question Answering via Sparse Transformer Matching Retrieval}. In \bibinfo{booktitle}{\emph{Proceedings of the 2021 Conference of the North American Chapter of the Association for Computational Linguistics: Human Language Technologies}}, \bibfield{editor}{\bibinfo{person}{Kristina Toutanova}, \bibinfo{person}{Anna Rumshisky}, \bibinfo{person}{Luke Zettlemoyer}, \bibinfo{person}{Dilek Hakkani-Tur}, \bibinfo{person}{Iz~Beltagy}, \bibinfo{person}{Steven Bethard}, \bibinfo{person}{Ryan Cotterell}, \bibinfo{person}{Tanmoy Chakraborty}, {and} \bibinfo{person}{Yichao Zhou}} (Eds.). \bibinfo{publisher}{Association for Computational Linguistics}, \bibinfo{address}{Online}, \bibinfo{pages}{565--575}.
\newblock
\href{https://doi.org/10.18653/v1/2021.naacl-main.47}{doi:\nolinkurl{10.18653/v1/2021.naacl-main.47}}


\end{thebibliography}

\clearpage
\appendix
\section{Problem Statement}\label{app:PB}
In product search, the goal of the first-stage retrieval is to recall as many products relevant to the user query as possible to serve as the candidate set for the subsequent ranking stage.Formally, the first-stage retrieval task in product search is defined as follows: Given a query set $\mathcal{Q} = \{q_1, q_2, \ldots, q_m\}$ and a product database $\mathcal{I} = \{d_1, d_2, \ldots, d_n\}$, where for each query $q_i$, the database contains a relevant product item set $\mathcal{I}_{q_i} \subseteq \mathcal{I}$. The retrieval model needs to return top-$k$ retrieved products $\mathcal{R}_i = \{d_{r_1}, d_{r_2}, \ldots, d_{r_k}\}$ from the product database $\mathcal{I}$ for each query $q_i$, with the objective of retrieving as many products as possible from the relevant product set $\mathcal{I}_{q_i}$ to achieve high recall. The average recall across the query set is defined as:
\begin{equation}
\text{Recall} = \frac{1}{|\mathcal{Q}|} \sum_{i=1}^{|\mathcal{Q}|} \frac{|\mathcal{R}_i \cap \mathcal{I}_{q_i}|}{|\mathcal{I}_{q_i}|},
\end{equation}
where $|\mathcal{R}_i \cap \mathcal{I}_{q_i}|$ represents the number of relevant products in the retrieval results for query $q_i$, and $|\mathcal{I}_{q_i}|$ represents the total number of relevant products for query $q_i$.

\vspace*{-1mm}
\section{Related Work}
\label{sec:related_work}
\heading{Dense retrieval} Leveraging the powerful 
representation capabilities of neural networks, dense retrieval models map text into dense embeddings to achieve superior retrieval performance, which has drawn broad attention from academia and industry \cite{guoSemanticModelsFirststage2022,devlin-etal-2019-bert,karpukhin-etal-2020-dense,gaoCondenserPretrainingArchitecture2021}. 
DPR \cite{karpukhin-etal-2020-dense} introduced a dual-encoder architecture with BERT \cite{devlin-etal-2019-bert}, enabling offline document encoding for efficient retrieval, while ColBERT \cite{10.1145/3397271.3401075} enhanced effectiveness via late interaction at the token level.
Further advances such as unsupervised pre-training \cite{izacard2022unsuperviseddenseinformationretrieval,gaoUnsupervisedCorpusAware2022,lei-etal-2023-unsupervised}, hard negative mining \cite{xiong2020approximatenearestneighbornegative}, and knowledge distillation \cite{hofstätter2021improvingefficientneuralranking,10.1145/3404835.3462891} have continued to improve the performance of dense retrieval models. 
Recently, researchers have begun exploring LLMs' potential for retrieval tasks \cite{wang-etal-2024-improving-text,
leeNVEmbedImprovedTechniques2025,taoLLMsAreAlso2024,liuLlama2VecUnsupervisedAdaptation2024,man-etal-2024-ullme}. 
However, LLMs use causal attention mechanisms, where each token can only attend to previous tokens in the sequence, which limits their ability to learn comprehensive text representations. 
To address this, LLM2Vec \cite{behnamghaderLLM2VecLargeLanguage2024} and Nv-Embed \cite{leeNVEmbedImprovedTechniques2025} introduced bidirectional attention with adaptive training, while Echo \cite{springerRepetitionImprovesLanguage2024} embedding duplicated inputs to expose tokens to full context.
However, dense embeddings suffer from complete black-box opacity and high index storage costs, posing challenges for large-scale industrial applications.

\heading{Sparse retrieval} 
Sparse retrieval \cite{9357643,10.1561/1500000019,10.1145/3404835.3463098} operates on explicit, term-level signals to compute relevance.
Classical models like BM25 \cite{10.1561/1500000019} rely on statistical term weighting and inverted indices, offering strong baselines due to their efficiency and robustness.
To capture richer semantics, learned sparse retrieval methods incorporate neural networks to assign more informative weights to terms \cite{10.1145/3269206.3271800,10.1145/3331184.3331303,dai2019contextawaresentencepassagetermimportance,10.1145/3397271.3401204,10.1145/3366423.3380258}.
However, they still depend heavily on literal term overlap. Given the significant length asymmetry between queries and documents, this leads to persistent vocabulary mismatch. 

To address this, some methods expanded terms in addition to weighting them \cite{nogueira2019documentexpansionqueryprediction,zhao-etal-2021-sparta,10.1145/3397271.3401262,nogueiraDocumentRankingPretrained2020}, but their performance lagged behind dense models. 
SPLADE \cite{10.1145/3404835.3463098}, building on SparTerm \cite{baiSparTermLearningTermbased2020}, optimize the pooling strategy for term weights and introduce FLOPS loss \cite{paria2020minimizingflopslearnefficient} for sparsification. 
Later versions added hard negatives and distillation, achieving dense-level performance in passage retrieval \cite{formalSPLADEV2Sparse2021,10.1145/3477495.3531857}, with follow-up work exploring fine-grained query-document interactions \cite{10.1145/3539618.3592065,10.1145/3539618.3591977,kongSparseEmbedLearningSparse2023}. 
Inspired by SPLADE and recent LLM-based dense retrieval, researchers have begun adapting LLMs for sparse retrieval. 
\citet{zengScalingSparseDense2025} studied the scaling laws of sparse retrievers based on LLMs, while CSPLADE \cite{xuCSPLADELearnedSparse2025}, building on Nv-Embed \cite{leeNVEmbedImprovedTechniques2025} and echo embedding \cite{springerRepetitionImprovesLanguage2024}, tackled attention-related challenges in applying SPLADE  \cite{10.1145/3404835.3463098} to LLMs. 
However, beyond replacing the model backbone, the core methodologies of these studies remain similar to SPLADE, and their generalization to the product search domain has not been confirmed.

\section{Experimental Setup Details}
\label{sec:appendix_exp_setup}

\subsection{Dataset Details}
\label{sec:appendix_dataset_details}
Details about our evaluation dataset are as follows:
\begin{itemize}[leftmargin=*]
\item \textbf{Multi-CPR E-commerce} \cite{10.1145/3477495.3531736}. Multi-CPR is a publicly available, multi-domain Chinese passage retrieval dataset. For our experiments, we utilize its E-commerce subset, which is sourced from real-world search scenarios on Taobao. This subset contains a corpus of over 1 million passages, from which we use a training set of 100,000 query-passage pairs and a test set of 1,000 queries. Each query in both the training and test sets is paired with a single, human-annotated positive item, ensuring a one-to-one correspondence.

\item \textbf{Taobao-Internal}. To further validate our approach in a real-world industrial setting, we construct a new dataset by sampling approximately 1.07 million query-item pairs from the real user click logs of Taobao Search in June 2025. The user clicks serve as the ground truth for relevance. A single query in this dataset may correspond to 1 to 10 clicked items. We create a training set of 270,000 query-item pairs and a test set of 1,000 queries along with their associated clicked items. 
\end{itemize}
The detailed token length statistics after tokenization by the Qwen-2.5-3B tokenizer are provided in Table~\ref{tab:dataset_stats_full}.
\begin{table}[h]
\centering
\caption{Full token length statistics of the datasets.}
\label{tab:dataset_stats_full}
\begin{tabular}{@{}l ccc c ccc@{}}
\toprule
\multirow{2}{*}{\textbf{Dataset}} & \multicolumn{3}{c}{\textbf{Query Length}} & \phantom{abc} & \multicolumn{3}{c}{\textbf{Item Length}} \\
\cmidrule{2-4} \cmidrule{6-8}
 & Min & Avg & Max & & Min & Avg & Max \\
\midrule
Multi-CPR & 1 & 5 & 23 & & 2 & 25 & 92 \\
Taobao-Internal & 2 & 6 & 25 & & 2 & 25 & 74 \\
\bottomrule
\end{tabular}
\end{table}
As the table shows, the average lengths of queries and items are short. This is particularly notable when contrasted with the vocabulary size of our Qwen-2.5-3B backbone, which exceeds 150,000, highlighting the extreme sparsity of the task.
\subsection{Implementation details}
\label{sec:implement}
Our backbone model is Qwen2.5 \cite{qwen2025qwen25technicalreport}, and we experiment with its 1.5B, 3B, and 7B versions. Unless otherwise specified, the default backbone is Qwen2.5-3B. For the lexical focusing window, the default sizes are $k_q=256$ and $k_d=512$. The dynamic window variants use sizes of $(256, 128, 64)$ for queries and $(512, 256, 128)$ for items. The dynamic window strategy adaptively shrinks the window size during training: when more than 90\% of queries or items have activated dimensions fewer than the current window size, the window automatically contracts to the next smaller size in the sequence. This adaptive mechanism ensures that the lexical focusing window remains appropriately sized relative to the actual sparsity patterns observed in the data, preventing over-constraining when representations naturally become sparser during training. The maximum sequence length for queries and items is set to 64, except for PROSPER\textsubscript{echo-emb}, where it is 128 due to duplicate input. For our baselines, dense retrievers are trained using the Tevatron~\cite{gao2022tevatronefficientflexibletoolkit} library, with indexing and search handled by Faiss~\cite{8733051}. Sparse retrievers use an internal database in Taobao for storage and match. For BM25 baselines, we use standard hyperparameters: $k_1=1.2$, $b=0.75$, and smoothing parameter $\delta=0.25$.

We train all models for five epochs on both datasets using eight NVIDIA H20 96GB GPUs. For each dataset, we randomly sample 500 queries from the training set to form a validation set, which is used to evaluate model performance after each epoch. The final model is selected based on the best performance on the validation set. The learning rate is set to $3e^{-5}$ with a linear warm-up for about 0.3 epochs and a batch size per device of 64. We use the AdamW optimizer \cite{loshchilov2019decoupledweightdecayregularization} with a weight decay of 0.1. The FLOPS regularization parameters, $\lambda_q$ and $\lambda_d$, are also quadratically increased to their target values of $5e^{-3}$ and $1e^{-3}$, respectively (about 1.5 epochs in our experiments).

\section{Supplementary experiments}
\subsection{Sparsification Strategy Analysis}
\label{sec:sparsity_analysis}
\begin{figure}[h]
    \centering
    \includegraphics[width=\linewidth]{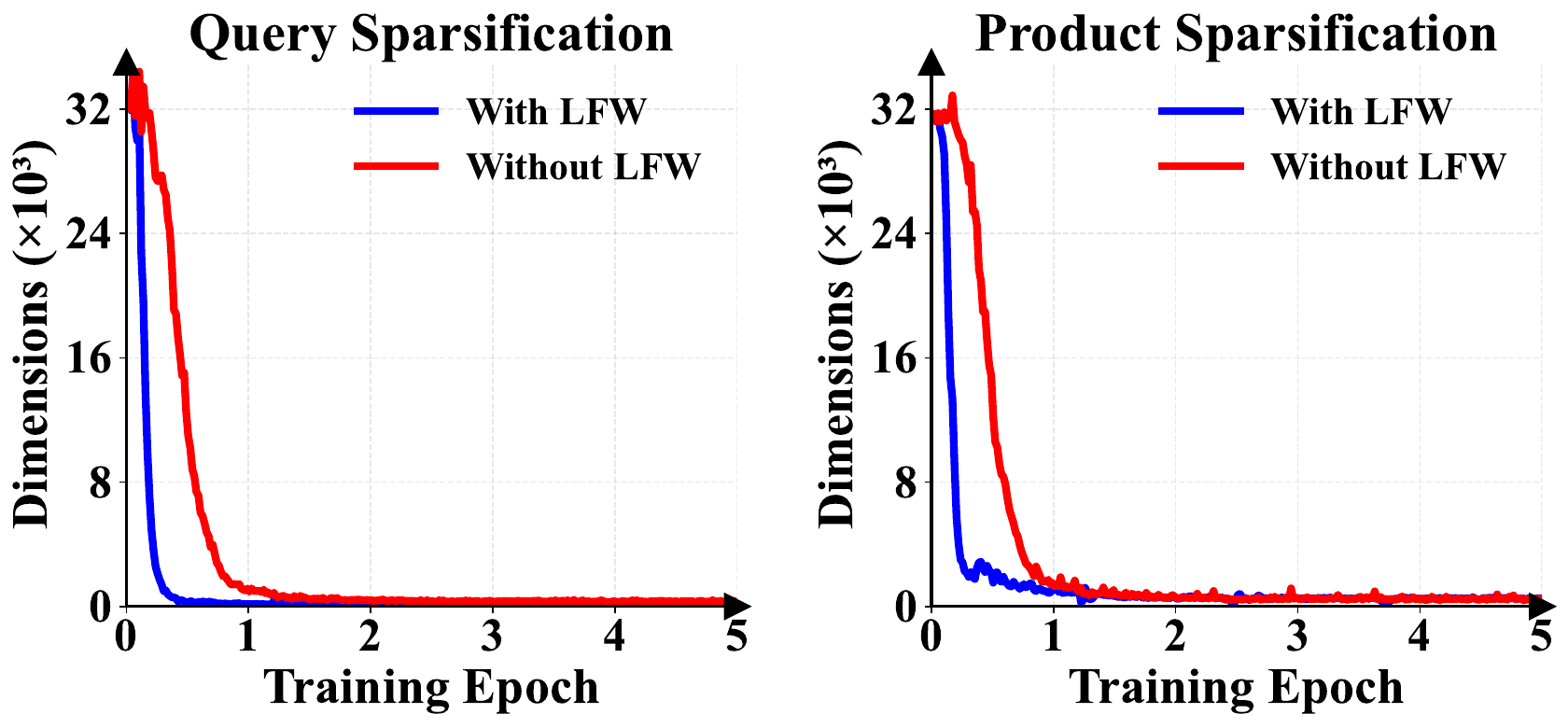}
    \caption{Impact of lexical focusing window on sparsification during training. The figure shows how the number of activated dimensions in query and item representations evolve with and without the LFW.}
    \label{fig:sparsity_comparison}
\end{figure}

Figure \ref{fig:sparsity_comparison} provides detailed analysis of how LFW affects the sparsification process during training, showing that LFW enables rapid dimension reduction in early training stages while achieving faster stabilization compared to models without LFW. The coarse-to-fine approach allows FLOPS regularization to handle fine-grained adjustments after the initial dimension reduction, resulting in more focused sparse representations that maintain high semantic quality while reducing computational overhead.

\subsection{Impact of normalization}
\label{sec:normalization}
To analyze the impact of normalization, we test the following:
\begin{enumerate*}[label=(\roman*)]
\item \textbf{PROSPER}\textsubscript{all-norm}, which applies $\ell_2$ normalization to both query and item vectors, making the scoring function equivalent to standard cosine similarity;
\item \textbf{PROSPER}\textsubscript{w/o-norm}, which removes all normalization, resulting in a standard dot product score;
\item \textbf{PROSPER}\textsubscript{d-norm}, which applies $\ell_2$ normalization only to the item vectors;
and \item \textbf{PROSPER}\textsubscript{$\ell_1$-norm}, which applies $\ell_1$ normalization to the query vector to explore its effect on term weighting.
\end{enumerate*}

Table \ref{tab:norm_impact} reveals the critical role of similarity functions in product search. Product search requires an asymmetric approach: queries need precise term expression to capture user intent, while items need rich semantic coverage to match diverse queries. Our analysis shows that applying $\ell_2$ normalization only to the query representation perfectly achieves this balance by emphasizing the relative importance of the query terms while preserving the absolute weight of item terms. This asymmetric approach significantly outperforms all variants. The standard dot product (PROSPER\textsubscript{w/o-norm}) and cosine similarity (PROSPER\textsubscript{all-norm}) fail to address this fundamental asymmetry, while normalization of $\ell_1$ (PROSPER\textsubscript{$\ell_1$-norm}) causes over sparsification by forcing the weights to sum to 1. Interestingly, normalizing only item representations (PROSPER\textsubscript{d-norm}) also underperforms, confirming that preserving absolute item weights is essential.
\begin{table}[h]
\centering
\caption{Impact of different similarity functions on retrieval performance.}
\label{tab:norm_impact}
\setlength{\tabcolsep}{6pt}
\begin{tabular}{@{}l@{\hspace{6pt}}c@{\hspace{6pt}}c@{\hspace{6pt}}c@{\hspace{6pt}}c@{}}
\toprule
\textbf{Variant} & \textbf{Hit@1} & \textbf{Hit@10} & \textbf{Hit@100} & \textbf{Hit@1000} \\
\midrule
PROSPER & 25.3 & 50.7 & 78.1 & 93.9 \\
PROSPER\textsubscript{$\ell_1$-norm}   & 15.8 & 39.9 & 68.0 & 90.0 \\
PROSPER\textsubscript{w/o-norm}   & 17.0 & 40.9 & 68.0 & 90.0 \\
PROSPER\textsubscript{all-norm}  & 17.1 & 39.7 & 67.2 & 87.2 \\
PROSPER\textsubscript{d-norm}   & 16.1 & 42.3 & 71.7 & 91.1 \\
\bottomrule
\end{tabular}
\end{table}
\begin{figure}[h]
    \centering
    \includegraphics[width=\linewidth]{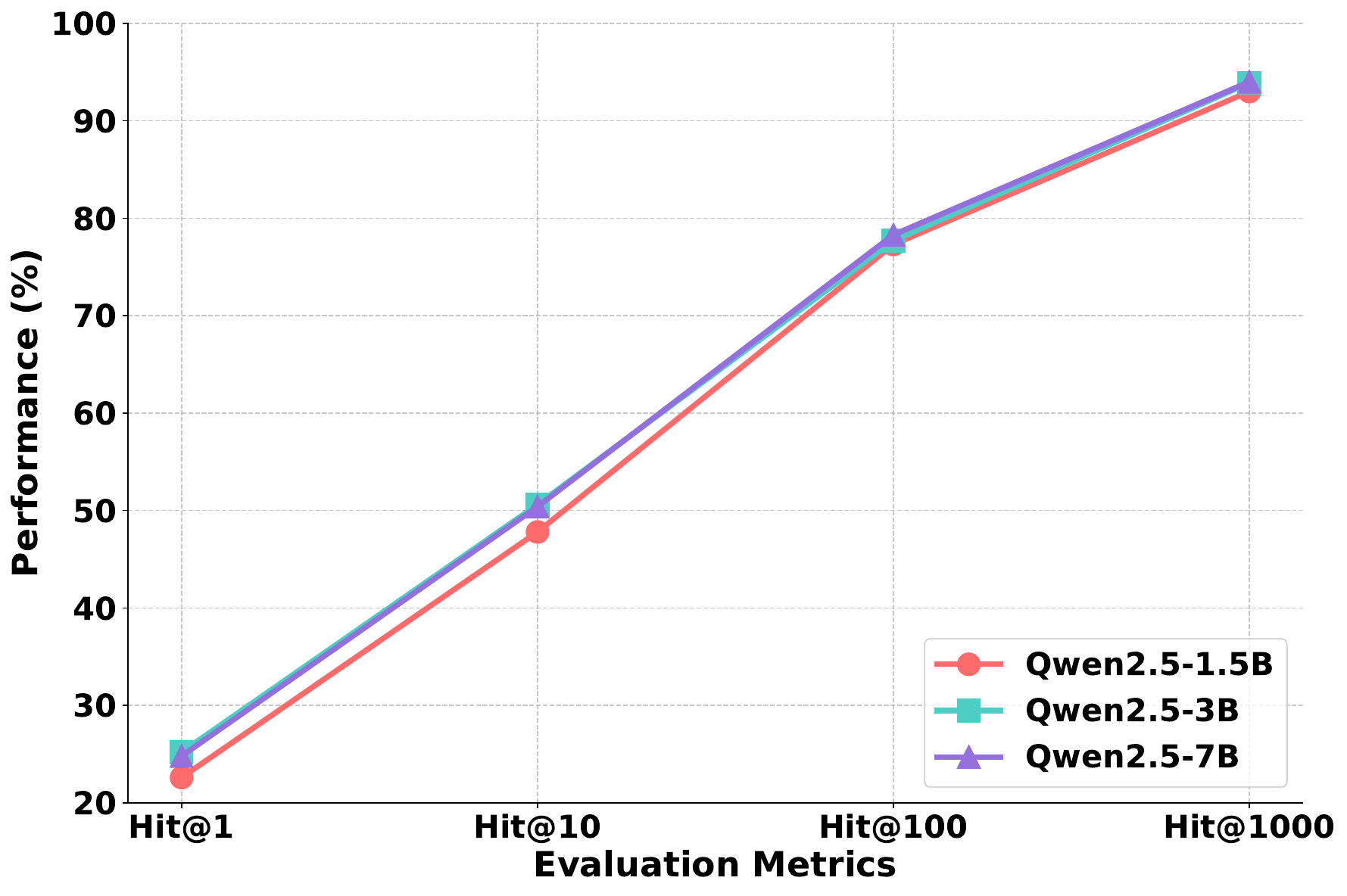}
    \caption{Performance comparison of PROSPER with different model sizes on the Multi-CPR test set.}
    \label{fig:parameter_scaling}
\end{figure}
\subsection{Parameter scaling}
\label{sec:para_scal}
To investigate the impact of model size, we experiment with PROSPER using Qwen2.5 backbones of three different scales: 1.5B, 3B, and 7B. As illustrated in Figure~\ref{fig:parameter_scaling}, we find that there is no obvious scaling law for our task. While larger models provide some gains, the performance on the crucial Hit@1000 recall metric is largely comparable across all scales. Therefore, for the subsequent online experiments, to achieve an optimal trade-off between effectiveness and deployment cost, we utilize the Qwen2.5-1.5B model as our backbone for large-scale online testing.

\section{Case Study}
\label{sec:case_study}
In this section, we present both offline and online case studies to demonstrate the effectiveness of PROSPER in real world scenarios. The offline case study analyzes the term expansion and weighting optimization results, while the online case study examines the exclusive recall results found in the multi-channel retrieval system in Taobao search.

\subsection{Offline case study}
\label{sec:offline_case_study}
To provide concrete insights into how PROSPER optimizes term expansion and weighting, we present detailed case studies showcasing the optimization results for the two examples mentioned in the preliminary section. Table~\ref{tab:case_study} demonstrates the before-and-after term expansion and weighting patterns for \begin{CJK}{UTF8}{gbsn}"爱立舍机油滤芯"\end{CJK} (Elysee oil filter) and \begin{CJK}{UTF8}{gbsn}"老捷豹副水箱"\end{CJK} (Old Jaguar water tank). The table illustrates how PROSPER effectively reduces lexical expansion hallucination while maintaining useful semantic expansions, leading to more focused and relevant term representations. Moreover, Table includes more comprehensive examples of expansion and weighting results.In addition, Table~\ref{tab:offline_case_study_2} also shows more offline cases.

\begin{table*}[htbp]
\centering
\caption{Case study of term expansion and weighting optimization by SP\textsubscript{Qwen-backbone} and PROSPER. The table shows the top-weighted terms before and after optimization for two representative product search queries. Terms are ranked by weight in descending order. Literal terms are shown in black, useful expansions in green, and noisy expansions in red.}
\label{tab:case_study}
\includegraphics[width=0.8\textwidth]{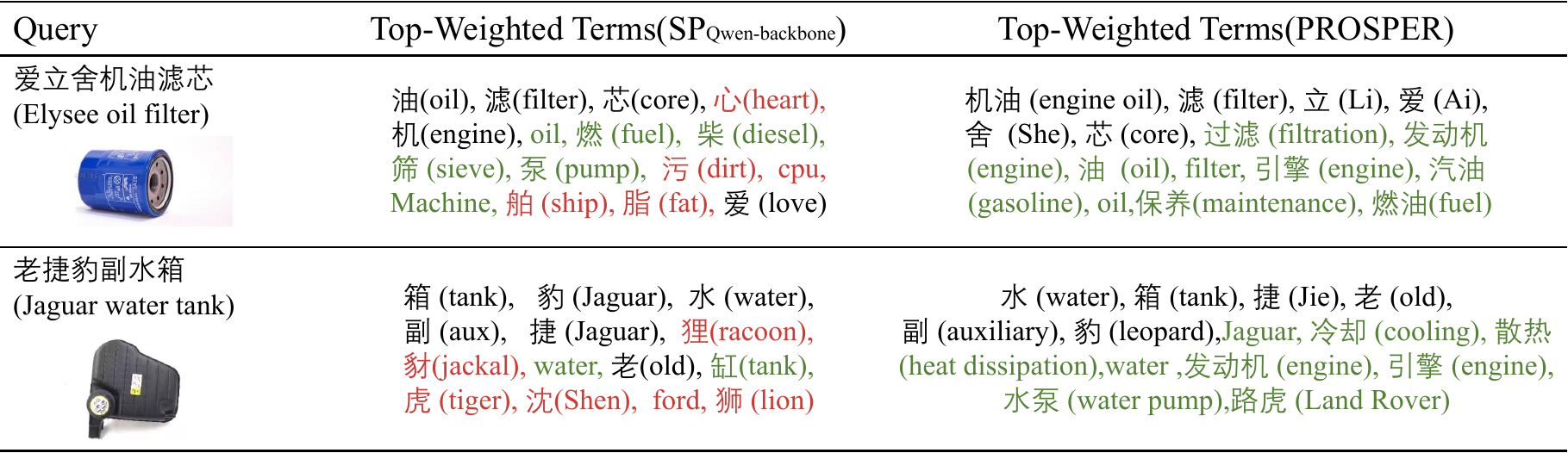}
\end{table*}

\begin{table*}[htbp]
\centering
\caption{More offline query and item cases of PROSPER.The table shows each term and its English translation, with weights in parentheses.Terms are ranked by weight in descending order. Literal terms are shown in black, useful expansions in green, and noisy expansions in red.}
\label{tab:offline_case_study_2}
\includegraphics[width=\textwidth]{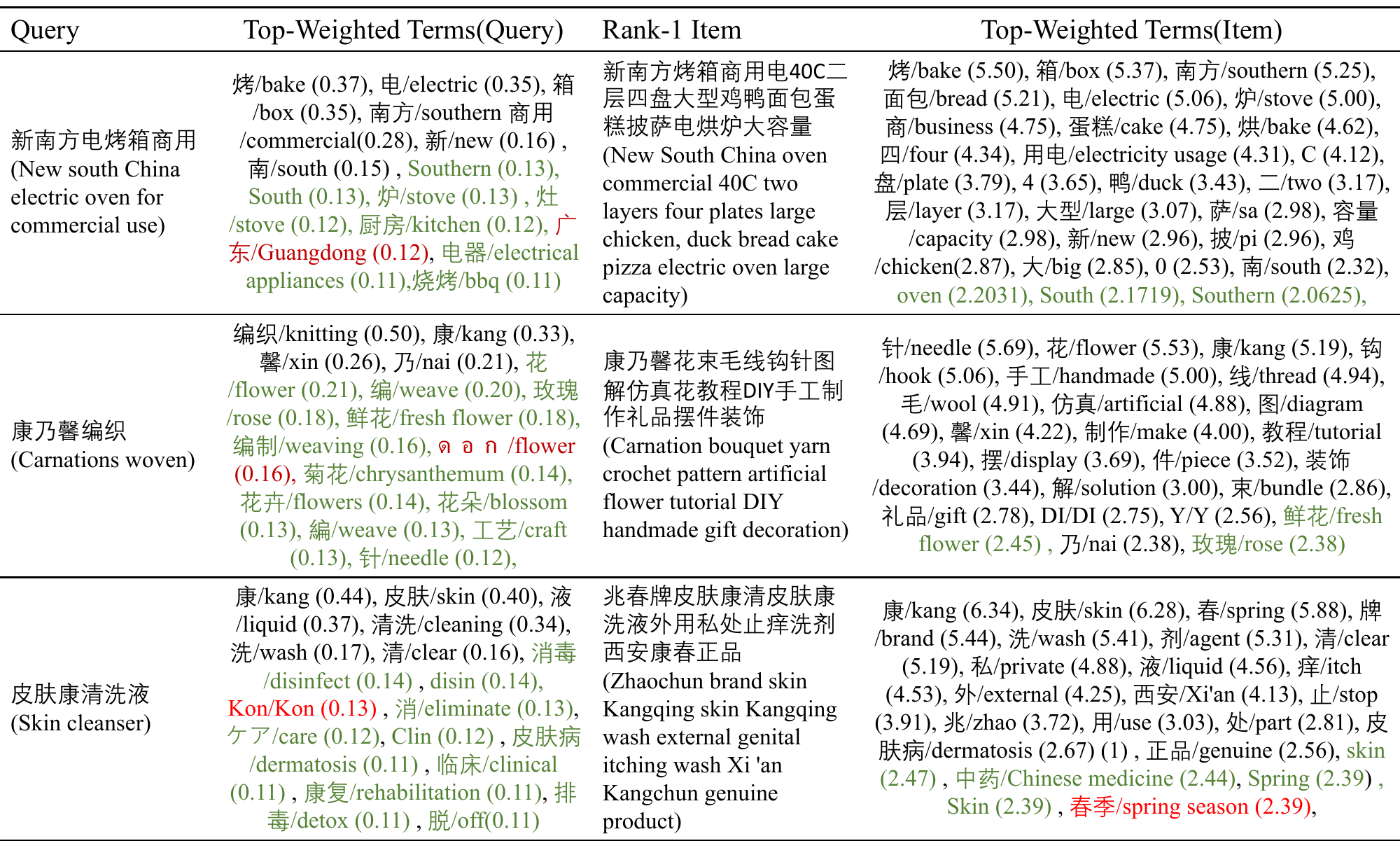}
\end{table*}

\subsection{Online case study}
\label{sec:online_case_study}
Our online case analysis focuses on products exclusively recalled by PROSPER, highlighting its unique contribution to the overall search performance. In this system, each retrieval channel is assigned a unique identifier, and PROSPER is designated as the eighth channel (with index 7). The retrieval source for each product is tracked using a bitmask called "recall\_types". A "recall\_types" value of $2^7=128$ indicates that the product was recalled solely by PROSPER. The relevance of recalled items is assessed by a Tabobao internal query-item relevance model, which assigns a "rnr" score: 2 for highly relevant, 1 for relevant, and 0 for irrelevant. As shown in Table~\ref{tab:online_case_study}, both "recall\_types" and "rnr" are annotated within the green box below each product in the table.  In practice, for each user query, there are multiple exclusive recall results by PROSPER as well as numerous multi-channel recall results. For convenience, in the table, we present one exclusive recall result by PROSPER and two recall results from other channels for each example query.

As illustrated by the cases in Table~\ref{tab:online_case_study}, even for common queries where other channels already demonstrate strong recall performance, they can still fail to retrieve some relevant products. PROSPER effectively addresses this gap by recalling these missing items. This ensures that the initial retrieval stage more comprehensively meets user needs, thereby enhancing the performance of the Taobao search engine and contributing to incremental online revenue for the platform.

\begin{table*}[t]
\centering
\caption{Online case study of exclusive product recalls by PROSPER in Taobao's hybrid retrieval system. The left column shows products exclusively recalled by PROSPER ("recall\_types" = 128), while the right column shows products for the same queries recalled by other channels. The "rnr" score indicates the relevance level.}
\label{tab:online_case_study}
\includegraphics[width=\textwidth]{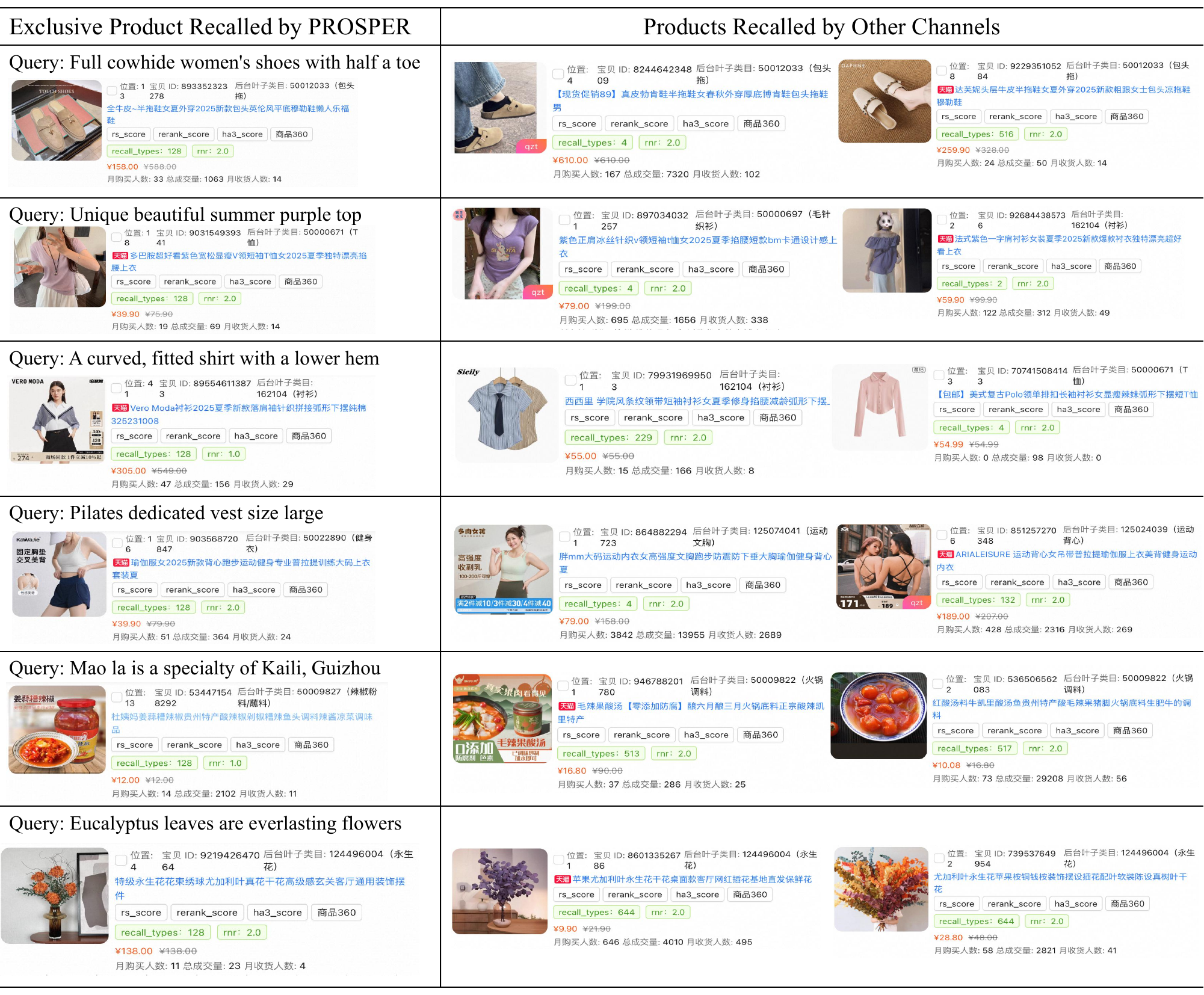}
\end{table*}

\end{document}